\documentclass[%
 reprint,
 amsmath,amssymb,
 aps,
]{revtex4-1}
\mathchardef\mhyphen="2D
\usepackage{graphicx}
\usepackage{dcolumn}
\usepackage{bm}
\usepackage{float}
\usepackage[dvipsnames]{xcolor}

\usepackage{amssymb}   
\usepackage{feynmf}
\usepackage{slashed}
\usepackage{multirow}   
\usepackage{tikz}
\usepackage{color}

\begin{document}
\preprint{APS/123-QED}
\title{Learning to Identify Electrons}

\newcommand{\JH}[1]{{\color{black!40!magenta} [\textbf{Jessica:} {#1}]}}
\newcommand{\TT}[1]{{\color{black!40!orange} [\textbf{Tony:} {#1}]}}
\newcommand{\JC}[1]{{\color{black!40!green} [\textbf{Julian:} {#1}]}}
\newcommand{\dw}[1]{\textbf{\color{blue} (#1 --dw)}}
\newcommand{\gxPlot}[3]{
\begin{gathered}\includegraphics[width=#1\textwidth]{images/#2-pass#3-graph.pdf}\end{gathered}
}
\author{Julian Collado$^{a}$}
\author{Jessica N. Howard$^{b}$}
\author{Taylor Faucett$^{b}$}
\author{Tony Tong$^{b}$}
\author{Pierre Baldi$^{a}$}
\author{Daniel Whiteson$^{b}$}
 \email{Corresponding author: pfbaldi@ics.uci.edu}

\affiliation{$^{a}$Department of Computer Science, University of California, Irvine, CA, USA 92627}
\affiliation{$^{b}$Department of Physics and Astronomy, University of California, Irvine, CA, USA 92627}


    \begin{abstract}
    We investigate whether state-of-the-art classification features commonly used to distinguish electrons from jet backgrounds in collider experiments are overlooking valuable information. A deep convolutional neural network analysis of electromagnetic and hadronic calorimeter deposits is compared to the performance of typical features, revealing a $\approx 5\%$ gap which indicates that these lower-level data do contain untapped classification power. To reveal the nature of this unused information, we use a recently developed technique to map the deep network into a space of physically interpretable observables.  We identify two  simple calorimeter observables which are not typically used for electron identification, but which mimic the decisions of the convolutional network and nearly close the performance gap.
      \end{abstract}

\date{\today}

\maketitle


\section{Introduction} \label{sec:intro}

Production of electrons in high-energy collisions provides an essential handle on precision studies of the Standard Model~\cite{Aad:2016naf,Almeida:2018cld} as well as for searches for new physics~\cite{ATLAS:2011ad,Chatrchyan:2013iqa}. The identification of electrons, and their separation from backgrounds which mimic their signature, is therefore a critical element in the data analysis toolkit, especially at lower transverse momentum, where the backgrounds rise rapidly~\cite{Hoenig:2014dsa}.

In collider experiments, electrons are identified by an isolated track which aligns with a localized energy deposition, primarily in the electromagnetic calorimeter. The primary source of backgrounds is the production of hadronic jets, which typically feature multiple tracks and extended energy deposition in both electromagnetic and hadronic calorimeters, but can fluctuate to mimic electrons. The tracker and calorimeters, however, are very finely segmented, producing high-dimensional data which is difficult to analyze directly. A mature literature~\cite{Khachatryan:2015hwa,ATLAS-CONF-2016-024} contains higher-level features designed by physicists to highlight the distinct signature of the electron and suppress the backgrounds. The higher-level features define a lower-dimensional feature space.
 
Recent strides in machine learning for physics, particularly the advent of deep learning~\cite{Baldi:2014kfa,Guest:2016iqz,baldi2020deep} and image-processing techniques~\cite{Cogan:2014oua, Baldi:2016fql, deOliveira:2015xxd,deOliveira:2018lqd}, have demonstrated that high-level features designed by domain experts may not always fully capture the information available in the lower-level high-dimensional data. Specifically, the rich but subtle structure of the deposition of energy by jets provides a powerful potential handle for discrimination. Given their role as the dominant background, this suggests that additional classification power may be gained by applying image-based deep learning techniques to electrons.
 
In this study, we apply deep convolutional neural networks (CNNs) to the task of distinguishing between electrons and jets, using separate images from the electromagnetic and hadronic calorimeters.  Due to the black-box nature of their operation, we do not  propose to use CNNs in place of the high-level features. Instead, we apply CNNs as probe  the information content of the low-level data in comparison to the high-level features. We show that the classification performance of the image-based CNNs exceeds the performance of the high-level features in common use by Large Hadron Collider (LHC) experiments, by a small, but significant, margin.  We then  identify the source of the untapped information and construct novel high-level features that capture it.

This paper is organized as follows. In Section~\ref{sec:overview}, we outline our approach. In Section~\ref{sec:dataset}, we discuss the details of our image generation process and the corresponding dataset used for CNN experiments. In Section~\ref{sec:currentHLVars}, we review the existing state-of-the-art ATLAS and CMS high-level features, which we combine to derive our benchmark performance. In Section~\ref{sec:nn}, we provide details of neural network architectures and training. In Section~\ref{sec:perf}, we discuss the performance of these networks. In Section~\ref{sec:gap}, we search for new high-level features to bridge the gap between CNNs and standard features. In Section~\ref{sec:disc}, we summarize and discuss the results, providing an intuitive understanding of the underlying landscape. 
\section{Overview} 
\label{sec:overview}

This study explores whether low-level, high-dimensional, $\mathcal{O}(10^3)$, calorimeter data contains  information useful for distinguishing electrons from a major background not  captured by the standard suite of high-level features designed by physicists. Similar studies in jets or flavor tags have revealed such gaps~\cite{Guest:2016iqz,Baldi:2016fql}.

We probe this issue using a simulated dataset created with publicly available fast simulations tools~\cite{deFavereau:2013fsa}; while such samples do not typically match the fidelity of those generated with full simulations~\cite{Agostinelli:2002hh}, we refine the calorimeter description for this study and find the modeling sufficiently realistic for a proof-of-principle analysis. Our focus is on comparing physically motivated, high-level features to low-level image techniques on equal footing.  While we anticipate that the numerical results will be different when evaluated in a fully realistic scenario, the broad picture will likely remain the same. The technique described here is fairly general and applicable to more realistic experimental scenarios, so that valuable lessons can be learned in the present context.

We reproduce the standard suite of electron identification features, as described in Refs.~\cite{Khachatryan:2015hwa,ATLAS-CONF-2016-024}, in the context of our simulated description. We then compare their combined performance to that of deep convolutional neural networks (CNNs) which have been trained to analyze the lower-level calorimeter cells using image recognition techniques~\cite{Cogan:2014oua,deOliveira:2015xxd,Baldi:2016fql}. We do not advocate for the use of CNNs to replace high-level features whose designs are grounded in physics; CNNs are difficult to interpret and the low level and large dimensionality of the input  makes validation of the features and definition of systematic uncertainties nearly impossible. Instead, here we use the power of CNNs as a probe, to test whether further information is present in the low-level data. Having identified a gap, we then explore a complete space of novel high-level features, Energy Flow Polynomials (EFPs)~\cite{Komiske:2017aww} to interpret and bridge the gap.

\section{Dataset Generation} \label{sec:dataset}

In this section, we describe the process of generating simulated signal and background datasets, reproducing the standard suite of high-level features, and forming pixelated images from the electromagnetic and hadronic calorimeter deposits.

\subsection{Processes and Simulation} \label{subsec:processes}

Simulated samples of isolated electrons are generated from the production and electronic decay of a $Z'$ boson in hadronic collisions, $pp \rightarrow Z' \rightarrow e^+e^-$ at $\sqrt{s} = 13$ TeV.  We set $m_{Z'}$ to 20 GeV in order to efficiently produce electrons in the range $p_{\textrm{T}}=[10,30]$ GeV, where hadronic backgrounds are significant.  Simulated samples of background jets are generated via generic dijet production. Events were generated with {\sc MadGraph} v2.6.5 \cite{MadGraph},  decayed and showered with {\sc pythia} v8.235 \cite{Pythia}, with detector response described by  {\sc delphes} v3.4.1 \cite{deFavereau:2013fsa} using {\sc root} version 6.08\/00 \cite{ROOT}. 

Our configuration of {\sc delphes} approximates the ATLAS detector.  For this initial study, we model only the central layer of the calorimeters where most energy is deposited; future work will explore more detailed and realistic detector simulation.
However, we 
maintain the critical separation between the electromagnetic and hadronic calorimeters and their distinct segmentation. Our simulated electromagnetic calorimeter (ECal) has segmention of $(\Delta \phi, \Delta \eta) = (\frac{\pi}{126}, 0.025)$ while our simulated hadronic calorimeter (HCal) is coarser, $(\Delta \phi, \Delta \eta) = (\frac{\pi}{31}, 0.1)$.  This approach allows us to investigate whether information about the structure of the many-particle jet is useful for suppressing their contribution. See Ref~\cite{deOliveira:2018lqd} for an analysis of the information contained in the shape of shower  for individual particles.

 No pile-up simulation was included in the generated data, as pileup subtraction techniques have been shown to be effective~\cite{Berta:2014eza}. In total, we generated 107k signal and 107k background objects.

\subsection{ Electron Candidate Selection} \label{subsec:objsel}

\begin{figure}
    \centering
    \includegraphics[width=0.8\linewidth]{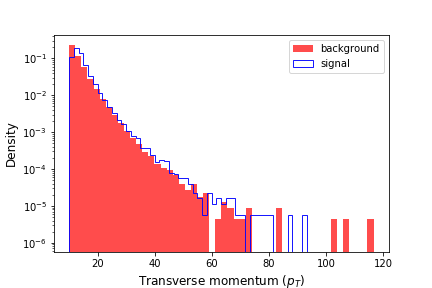}\\
    \includegraphics[width=0.8\linewidth]{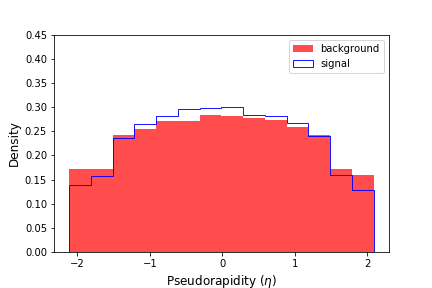}
    \caption{ Distribution of generated electron candidate $p_{\textrm{T}}$ and $\eta$ for simulated signal and background samples, before reweighting to match spectra. }
    \label{fig:pteta}
\end{figure}

We use {\sc delphes}' standard electron identification procedures where loose electron candidates are selected from charged particle tracks which align with energy deposits in the ECal.  We required object to have track $p_\textrm{T}> 10$ GeV and $|\eta|<2.125$, to avoid edge effects when forming calorimeter images, see Fig.~\ref{fig:pteta}.  For later training, the background objects are reweighted to match the $p_\textrm{T}$ distribution of the signal.

\subsection{ Image Formation} \label{subsec:imgform}

The cells of the calorimeter can be naturally organized as pixels of an image, allowing for use of powerful image-processing techniques. Each pixel contains the energy deposited in one cell. Alternatively, one may form images in which each cell represents $E_\textrm{T} = E / \cosh{\eta}$, which folds in the location of the object relative to the collision point.  For completeness  we initially consider images in which pixels represent $E$ and images where pixels represent $E_\textrm{T}$. Additionally, we create separate images for the ECal and HCal, in order to preserve the separate and powerful information they offer.  In total, four images are created for each electron candidate: ECal $E$, ECal $E_\textrm{T}$, HCal $E$, HCal $E_\textrm{T}$.

\begin{figure}[ht]%
    \centering
    Electromagnetic Calorimeter Images\\
    \includegraphics[width=0.45\linewidth]{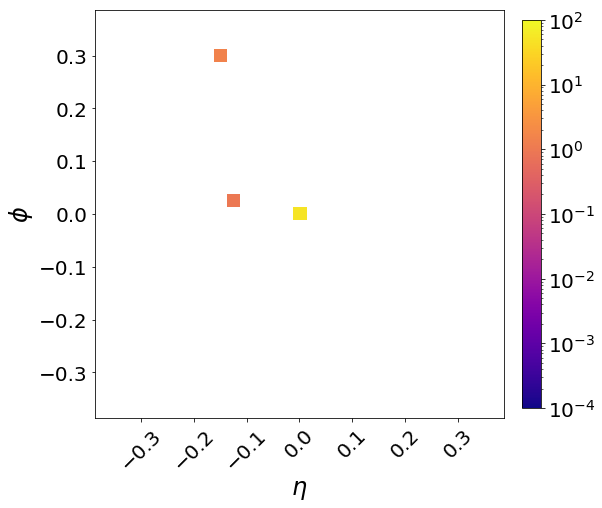}
    \label{fig:e_cal_sig_sample}
    \qquad
    \hspace{-2em}%
    \includegraphics[width=0.45\linewidth]{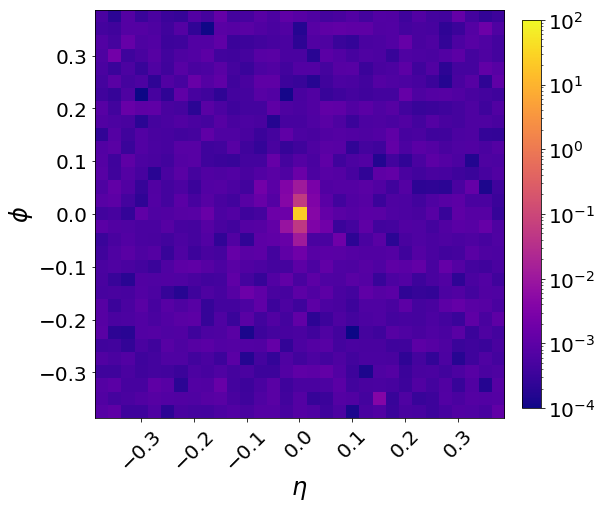}%
    \label{fig:e_cal_sig_avg}\\
    (a) Example Electron \ (b) Mean Electron\\
    \includegraphics[width=0.45\linewidth]{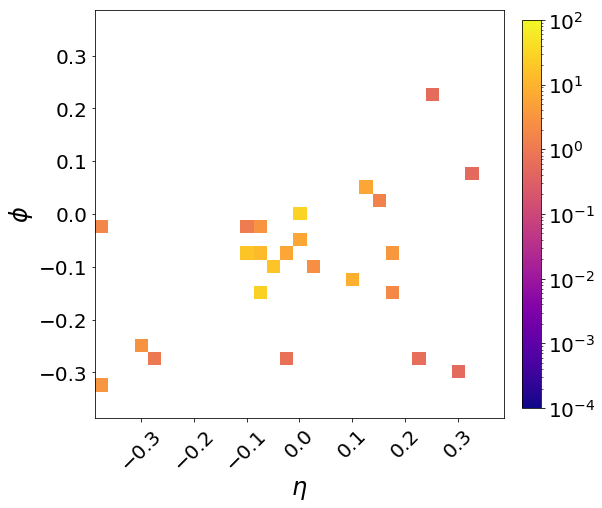}%
    \label{fig:e_cal_bg_sample}
    \qquad
    \hspace{-2em}%
    \includegraphics[width=0.45\linewidth]{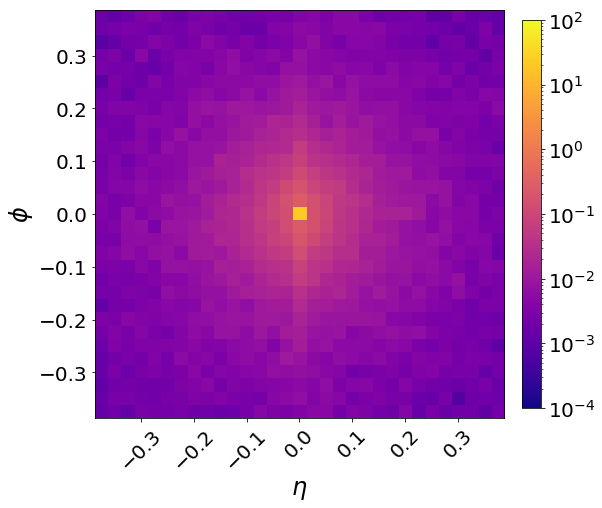}%
    \label{fig:e_cal_bg_avg}\\
    (c) Example Jet \ \ \ \ \ (d) Mean Jet\\
    \caption{ Images in the electromagnetic calorimeter for signal electrons (top) and background jets (bottom).  On the left are individual examples, on the right are mean images. See Fig.~\ref{fig:h_cal_img} for corresponding hadronic calorimeter images.}
    \label{fig:e_cal_img}
\end{figure}

\begin{figure}[ht]%
    \centering
    Hadronic Calorimeter Images\\
    \includegraphics[width=0.45\linewidth]{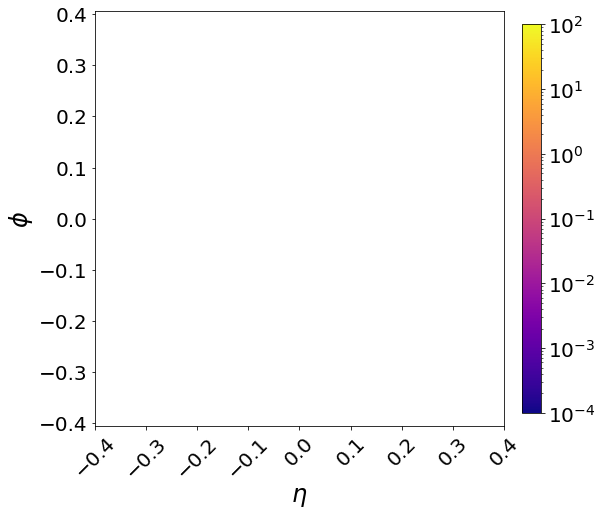}
    \label{fig:h_cal_sig_sample}
    \qquad
    \hspace{-2em}%
    \includegraphics[width=0.45\linewidth]{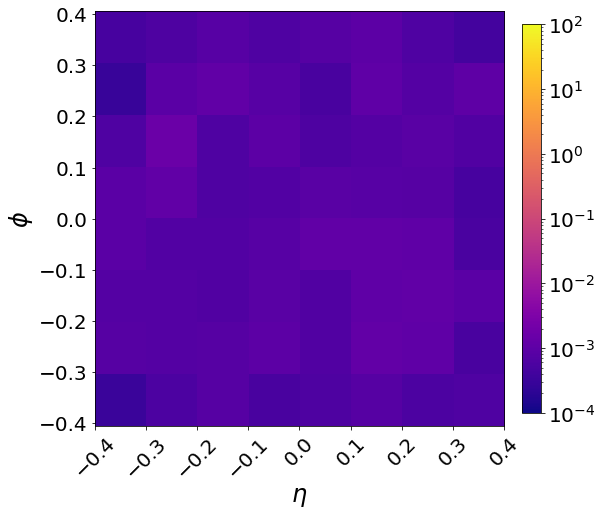}%
    \label{fig:h_cal_sig_avg}\\
    (a) Example Electron \ (b) Mean Electron\\
    \includegraphics[width=0.45\linewidth]{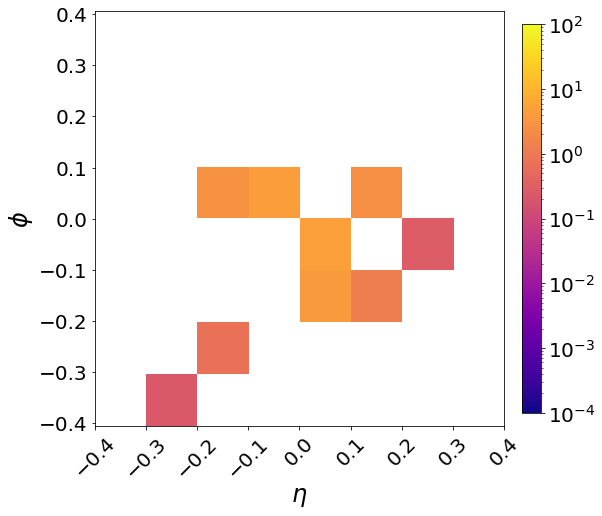}%
    \label{fig:h_cal_bg_sample}
    \qquad
    \hspace{-2em}%
    \includegraphics[width=0.45\linewidth]{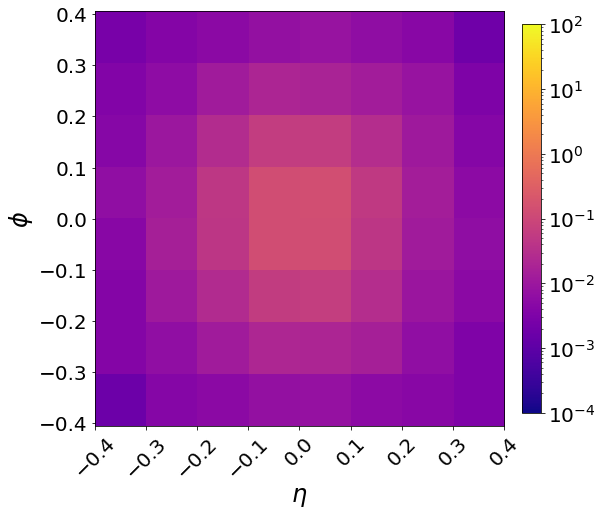}%
    \label{fig:h_cal_bg_avg}\\
        (c) Example Jet \ \ \ \ \ (d) Mean Jet\\
    \caption{ Images in the hadronic calorimeter for signal electrons (top) and background jets (bottom).  On the left are individual examples, on the right are mean images. See Fig.~\ref{fig:e_cal_img} for corresponding electromagnetic calorimeter images.}
    \label{fig:h_cal_img}
\end{figure}

The center of a calorimeter image is chosen to be the ECal cell with largest transverse energy deposit in the $9\times 9$ cell region surrounding the track of the highest $p_\textrm{T}$ electron in that event. This accounts for the curvature in the path of the electron as it propagates between the tracker and the calorimeter. The ECal image extends fifteen pixels in either direction, forming a $31\times 31$ image.  The HCal granularity is four times as course, and an $8\times 8$ image covers the same physical region. Figures~\ref{fig:e_cal_img} and~\ref{fig:h_cal_img} show example and mean images for the ECal and HCal, respectively.

\section{ Standard Classification Features} \label{sec:currentHLVars}

To assess the performance of the high-level classifications features typically used by ATLAS~\cite{ATLAS-CONF-2016-024} and CMS~\cite{Khachatryan:2015hwa} which identify electrons and reject jet backgrounds, we reproduce their form here, where relevant. 

Since electron candidates are confined to the longitudinal range $|\eta| < 2.125$, we only consider variables that are well-defined in this range. Additionally, we only consider variables which are based on information included in our simulation, to ensure the comparison uses information on equal footing. In addition, we do not perform clustering; where a feature calls for the cluster energy, we replace it with the total energy of the candidate image, a reasonable proxy for the cluster in our less-finely segmented simulation.  All high-level features are calculated from the ECal and HCal images, using $E$ or $E_\textrm{T}$ images where appropriate.


We reproduce seven features: $R_{\textrm{had}}$,  $\omega_{\eta 2}$,  $R_\phi$,  $R_\eta$, $\sigma_{\eta\eta}$, and two isolation quantities. Together, these capture the typical strategies of suppressing objects with significant hadronic energy or extended energy deposits. Definitions of each feature are below, and distributions for signal and background samples are shown in Fig.~\ref{fig:hl}.

\subsubsection{\textbf{Ratio of HCal and ECal  Energy: $R_\textrm{had}$}}

The feature $R_{had}$ relates the transverse energy ($E_\textrm{T}$) in the electromagnetic calorimeter to that in the hadronic calorimeter. Specifically,

\begin{align}
R_\textrm{had} &= \frac{\Sigma_i E^{\textrm{HCal}}_{\textrm{T,}i}}{\Sigma_j E^{\textrm{ECal}}_{\textrm{T,}j}}
\end{align}
where $i$ and $j$ run over the pixels in the HCal and ECal images, respectively. 

\subsubsection{\textbf{Lateral Width of the ECal Energy Shower:} $w_{\eta 2}$}
The lateral width of the shower in the ECal, $w_{\eta 2}$, is calculated as
\begin{align}
w_{\eta 2} = \sqrt{  \frac{\Sigma_i E_i (\eta_i)^2}{\Sigma_i E_i} -  (\frac{\Sigma_i E_i \eta_i}{\Sigma_i E_i})^2  }
\end{align}
where $E_i$ is the energy of the $i^{th}$ pixel in the ECal image and $\eta_i$ is the pseudorapidity of the $i^{th}$ pixel in the ECal image measured relative to the image's center. The sum is calculated within an  $(\eta \times \phi) = (3 \times 5)$ cell window centered on the image's center. 


\subsubsection{\textbf{Azimuthal and Longitudinal Energy Distributions: $R_{\phi}$ and $R_{\eta}$}}

To probe the distribution of energy in azimuthal ($\phi$) and longitudinal ($\eta$) directions, we calculate two features $R_\phi$ and $R_\eta$. Qualitatively, these relate the total ECal energy in a subset of cells to the energy in a larger subset of cells extended in either $\phi$ or $\eta$, respectively. Specifically,
\begin{align}
R_{\phi} = \frac{E_{3 \times 3}}{E_{3 \times 7}},\ \ \  R_{\eta} = \frac{E_{3 \times 7}}{E_{7 \times 7}}
\end{align}
where the subscript  indicates the number of cells included in the sum in $\eta$ and $\phi$ respectively. For example, $(\eta \times \phi) = (3 \times 7)$ is a subset of cells which extends $3$ cells in $\eta$ and $7$ in $\phi$ relative to the center of the image.


\subsection{\textbf{ Lateral Shower Extension: $\sigma_{\eta \eta}$}}

An alternative probe of the distribution of energy in $\eta$ is $\sigma_{\eta \eta}$. Specifically, 
\begin{align}
\sigma_{\eta \eta} = \sqrt{  \frac{\Sigma_i w_i (\eta_i - \bar{\eta})}{\Sigma_i w_i}}
\end{align}
Where  $w_i$ is the weighting factor $|\ln(E_i)|$ with $E_i$ being the ECal energy of the $i^{th}$ pixel. The sum runs over the non-zero cells in the $(\eta \times \phi) = (5 \times 5)$ subset of cells centered on the highest energy cell in the ECal. Here, $\eta_i$ is measured in units of cells away from center, $\bar{\eta}$, as $\eta_i \in 0$, $\pm 1$, or $\pm 2$ if we choose $\bar{\eta} = 0$.

\subsection{\textbf{ Isolation}}

Jets typically deposit significant energy surrounding the energetic core, where electrons are typically isolated in the calorimeter. To assess the degree of isolation, we sum the ECal energy in cells within the angular range $\Delta R = \sqrt{\Delta \eta^2 + \Delta \phi^2} < 0.3 $ or $ 0.4$, where $\Delta \eta$ and $\Delta \phi$ are measured from a given cell's center and the center of the image. 

\begin{figure}[ht]%
    \centering
    \includegraphics[width=0.45\linewidth]{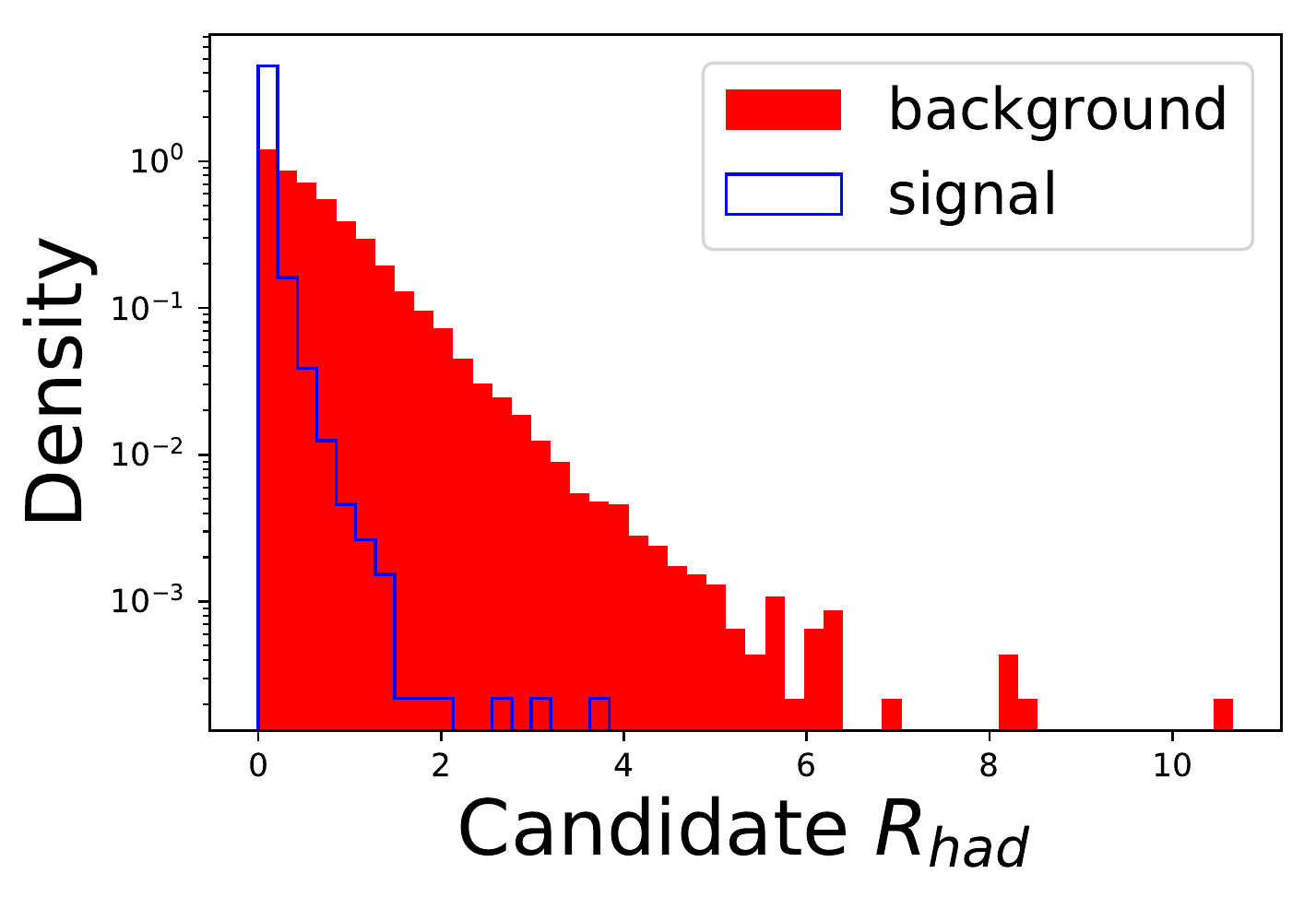}
    \label{fig:hl_distr_0}
    \hspace{-1em}%
    \qquad
    \includegraphics[width=0.45\linewidth]{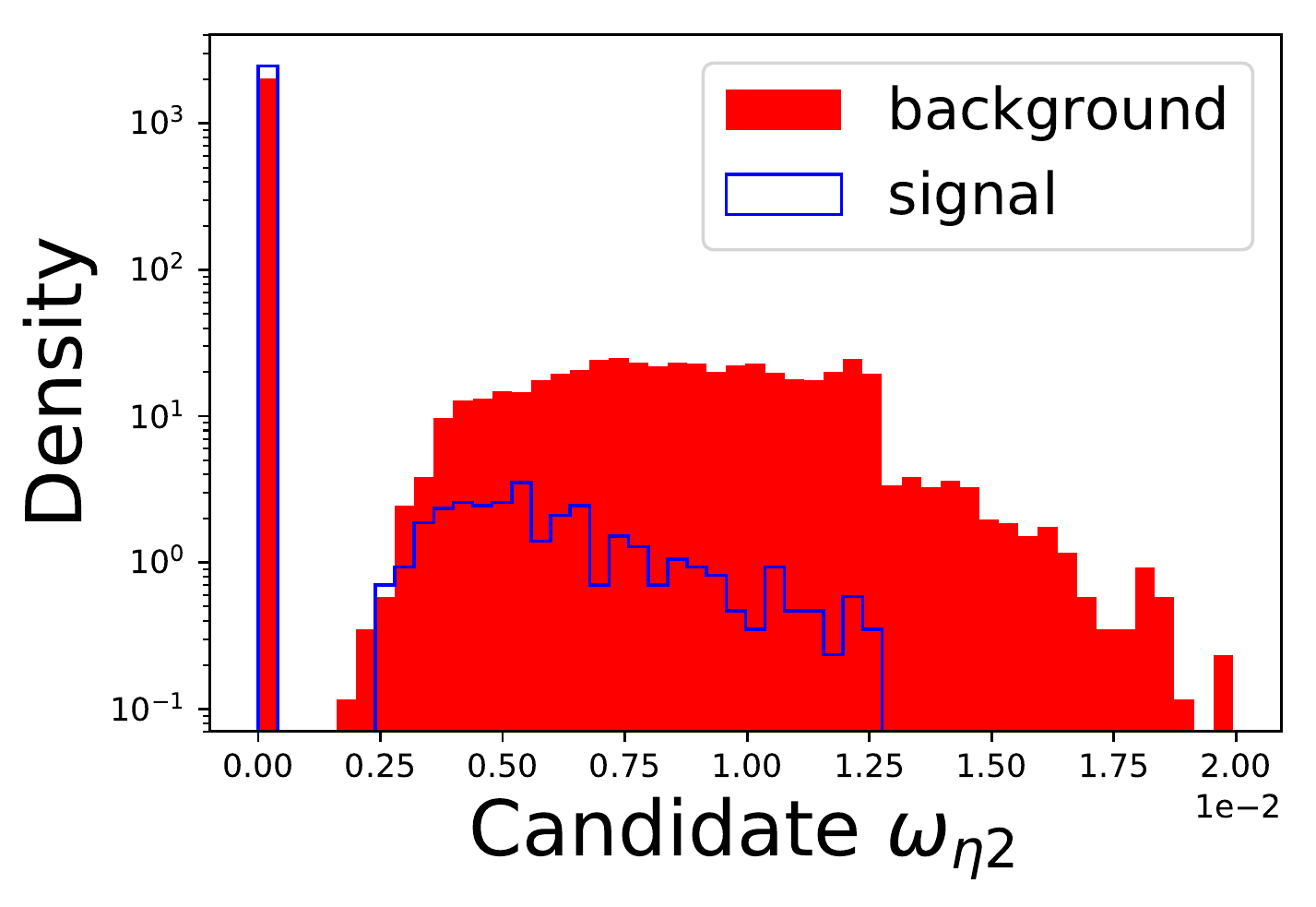}
    \label{fig:hl_distr_1}
    \\
    \includegraphics[width=0.45\linewidth]{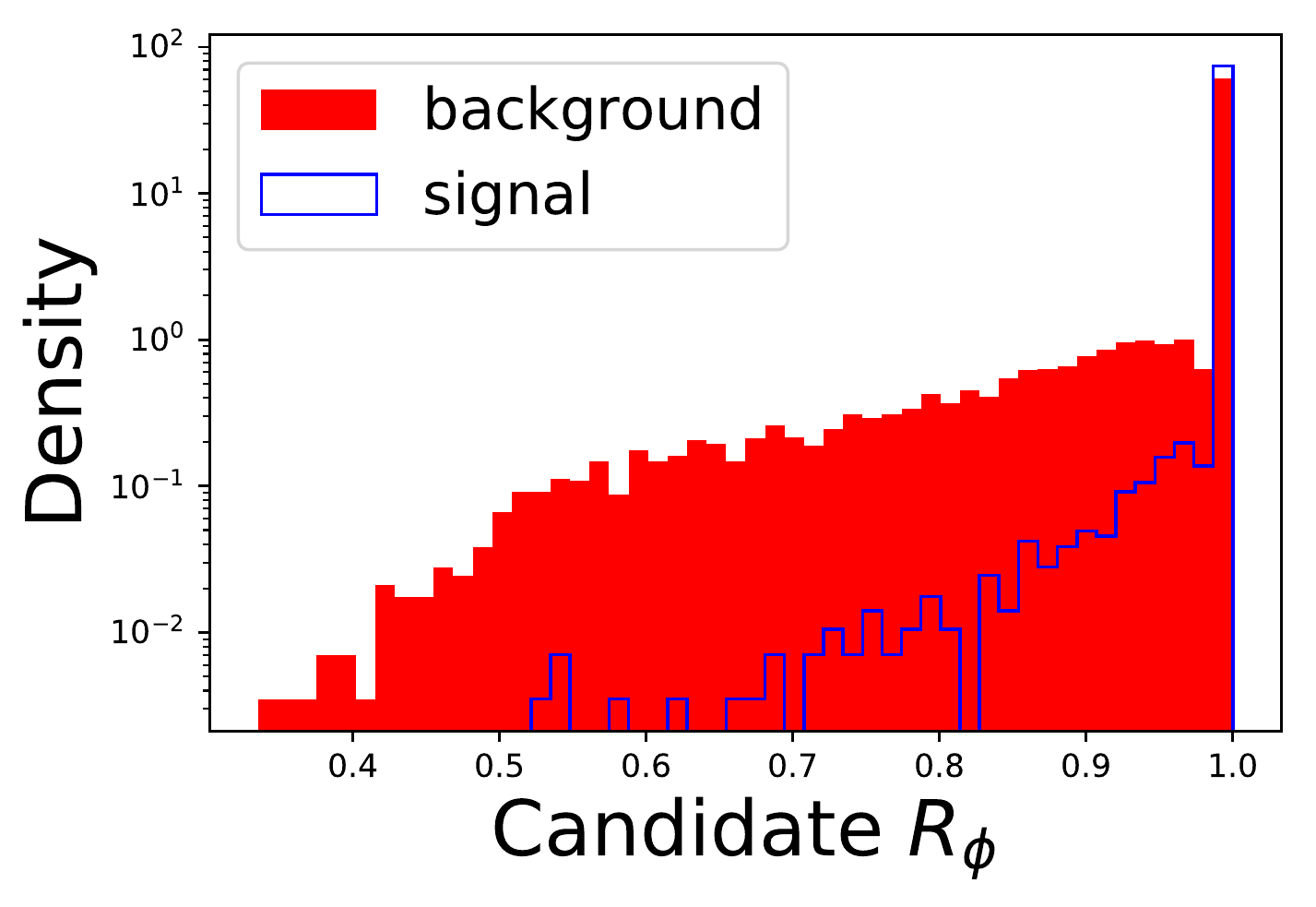}
    \label{fig:hl_distr_2}
    \hspace{-1em}%
    \qquad
    \includegraphics[width=0.45\linewidth]{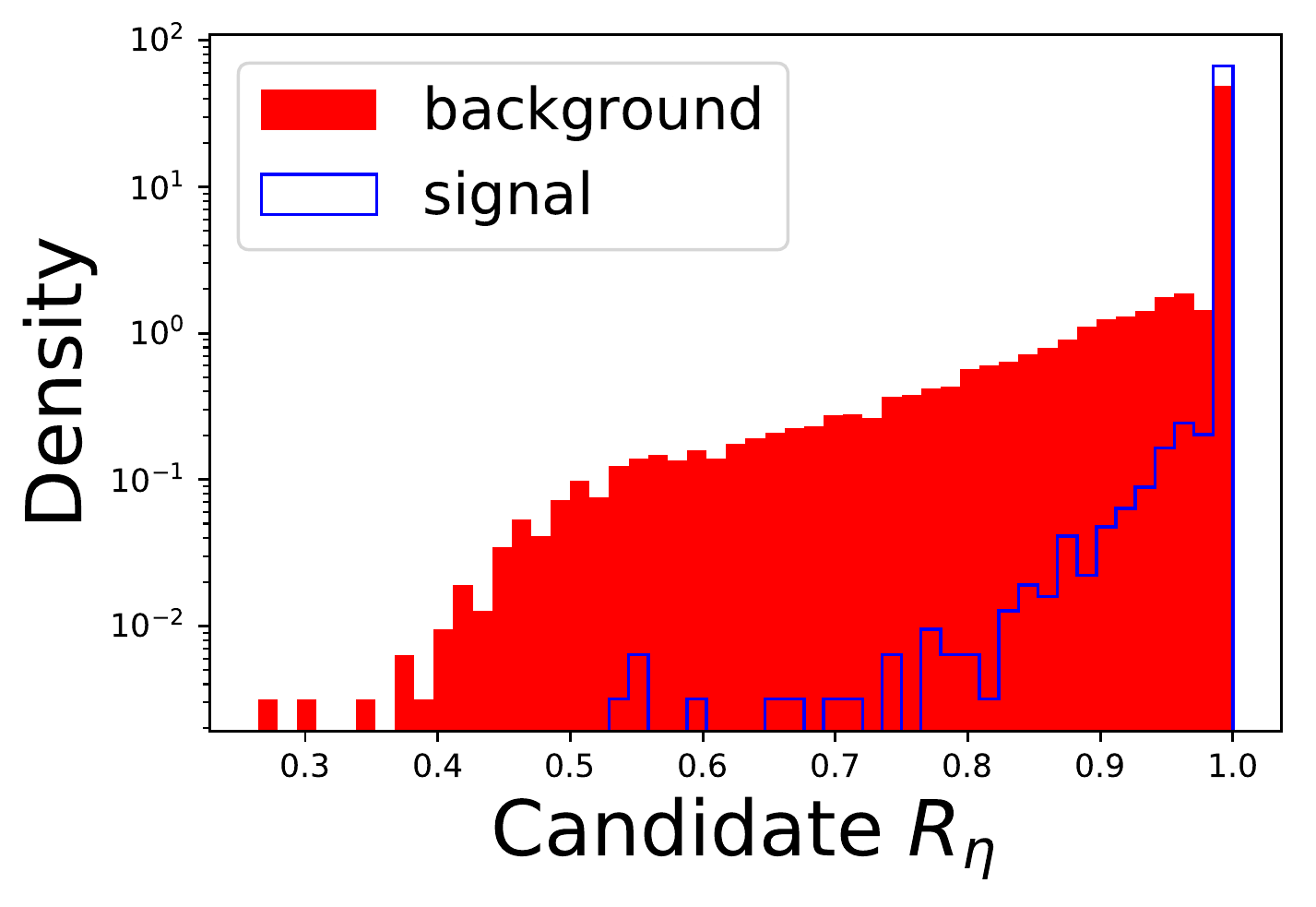}
    \label{fig:hl_distr_3}
    
    \includegraphics[width=0.45\linewidth]{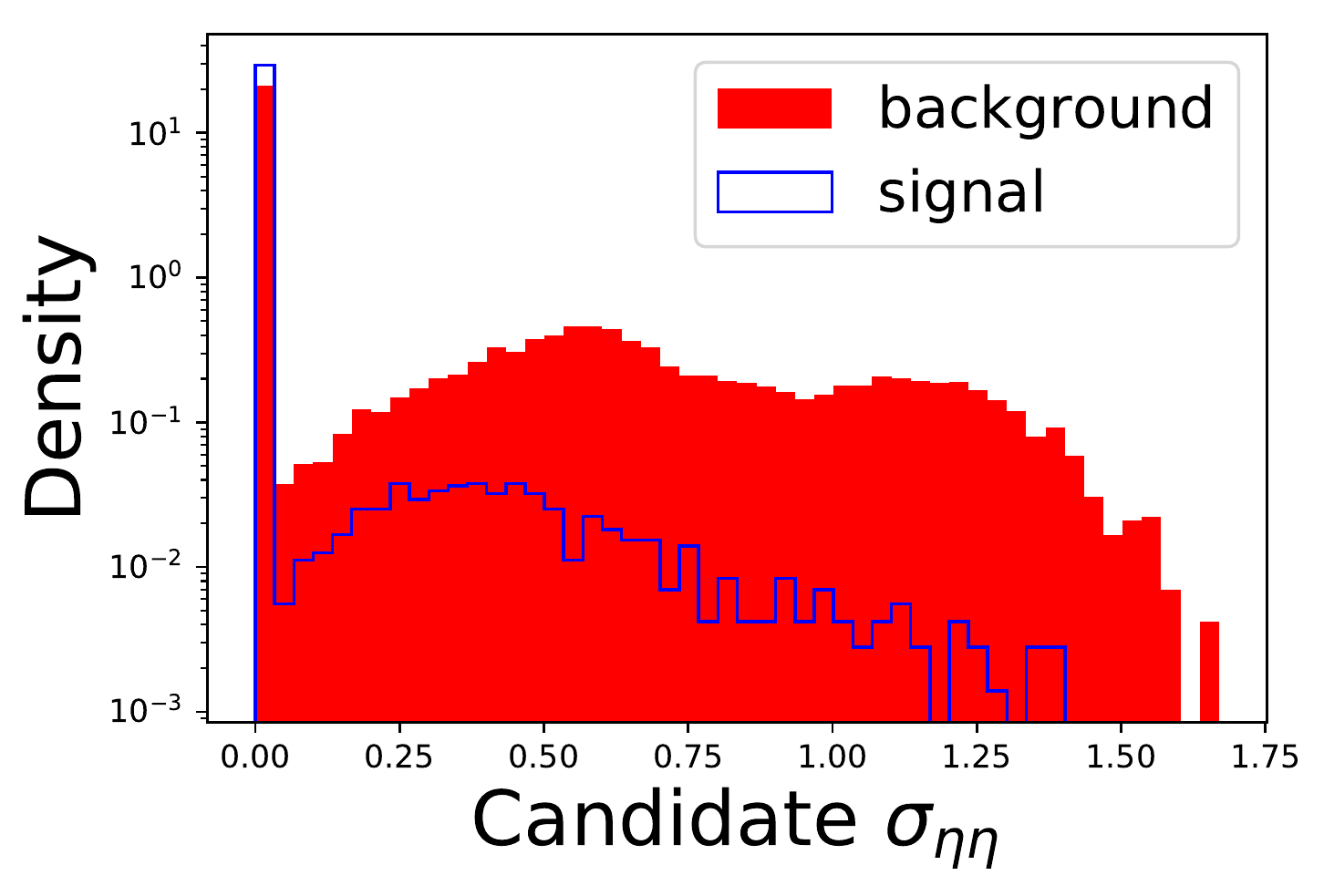}
    \label{fig:hl_distr_4}
    \hspace{-1em}%
    \qquad
    \includegraphics[width=0.45\linewidth]{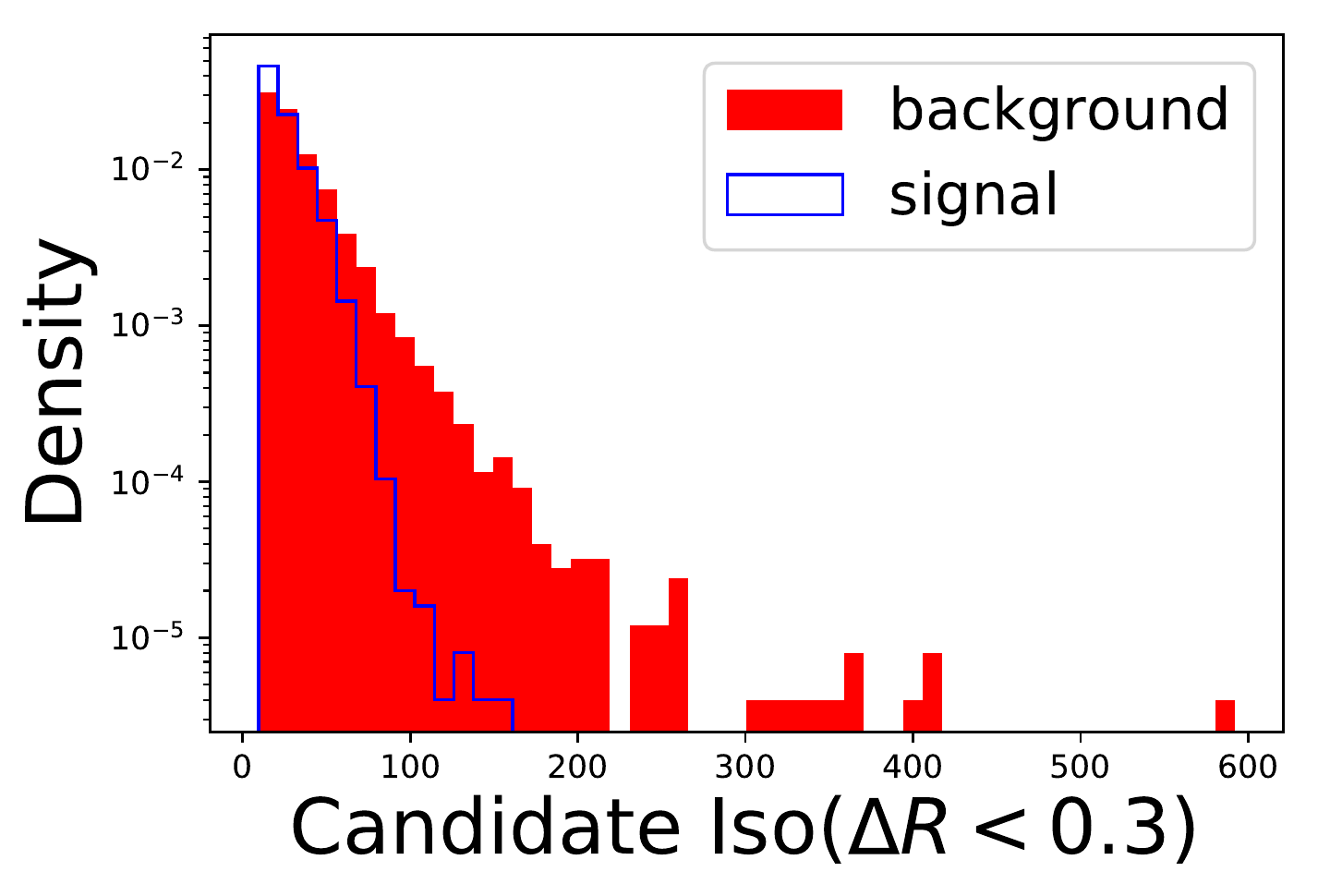}
    \label{fig:hl_distr_5}
    \\
    \includegraphics[width=0.45\linewidth]{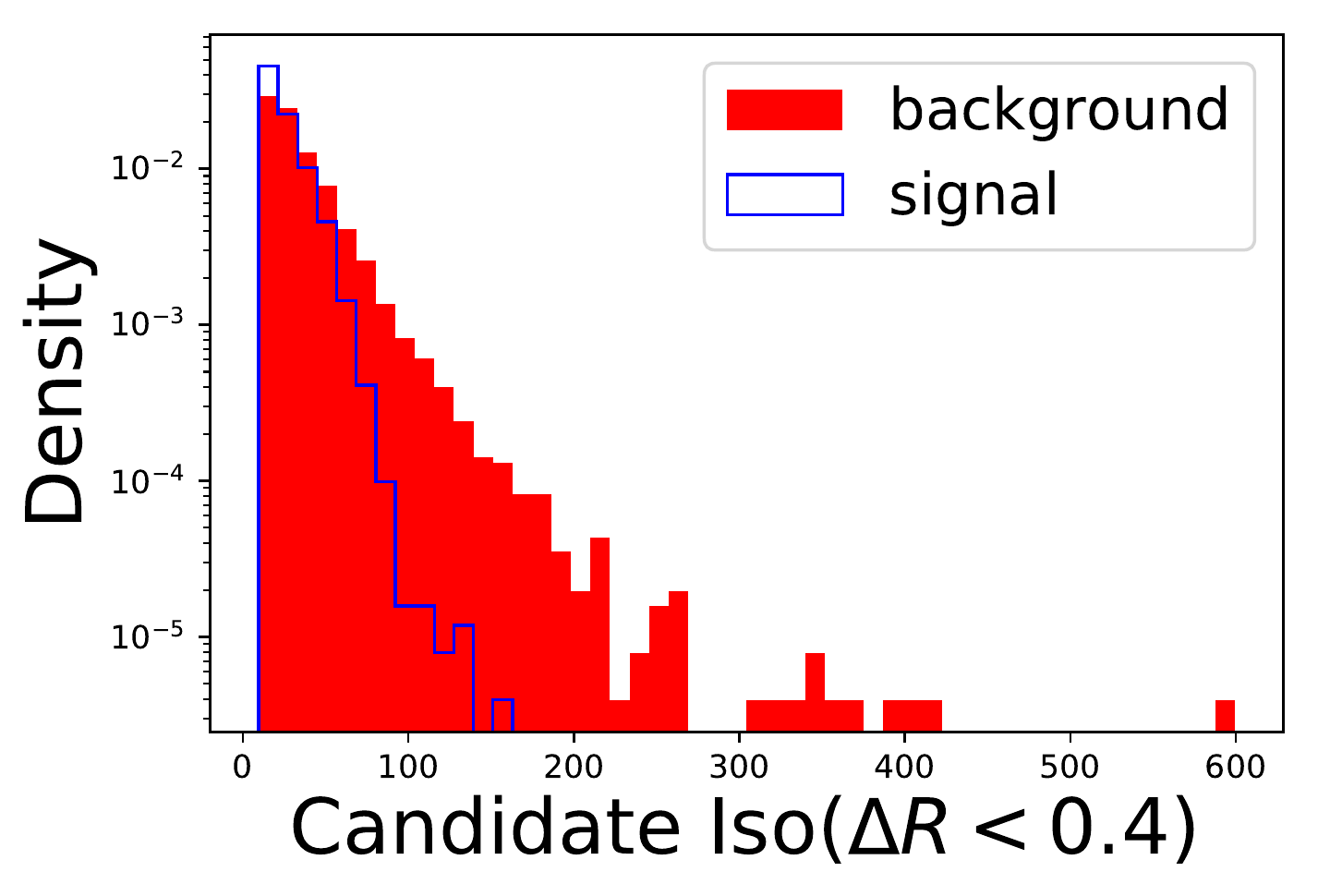}
    \label{fig:hl_distr_6}
        \hspace{-1em}%
    \qquad
    \includegraphics[width=0.45\linewidth]{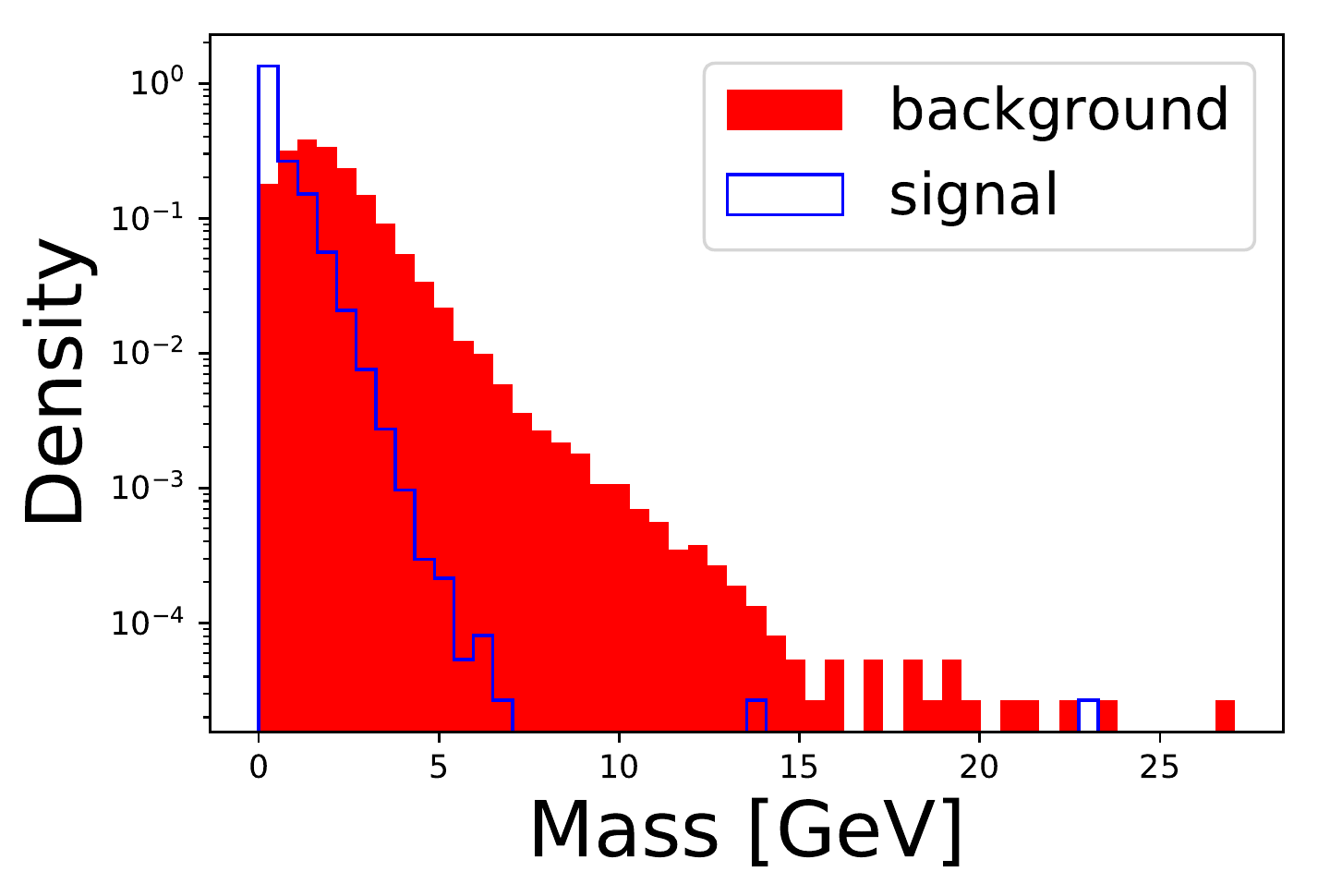}
    \label{fig:hl_distr_7}
    \caption{ Distribution of signal electron (red) and background jets (blue) for seven existing typically-used high-level features, as well as for mass. }
    \label{fig:hl}
\end{figure}

\section{ Neural Network Architectures and Training}
\label{sec:nn}


We construct multi-layer neural networks that accept low-level images, or high-level features, or both, with a sigmoidal logistic unit as their output unit to classify between signal and background.

Each image input is passed through a number of convolutional blocks, with each block consisting of two convolutional layers with $3 \times 3$ kernels,  rectified linear units \cite{RectifiedLinearUnits} as the activation function, and a final $2 \times 2$ maxpooling layer. Finally, the outputs of the maxpooling layer are flattened and concatenated with the high-level inputs to form a high-dimensional vector.
This high-dimensional vector is then processed by a sequence of fully connected layers with rectified linear units, using dropout\cite{Dropout,baldidropout14}.
The final output is produced by a single logistic unit and it can be interpreted as the probability of the input belonging to the signal class. The entire architecture is trained by stochastic gradient descent to minimize the relative entropy between the targets and the outputs, across all training examples.  

For each combination of high-level variables, we also train and tune multi-layer, fully connected, neural networks with a similar sigmoidal logistic unit at the top,

All models were implemented using {\sc{Keras}}~\cite{chollet2015keras} with {\sc{Tensorflow}}~\cite{tensorflow2015} as the backend and trained with a batch size of 128 with the {\sc{Adam}} optimizer~\cite{Adam2014}. The weights for all the models were initialized using orthogonal weights and each network was tuned using 150 iterations of bayesian optimizaton with the {\sc{Sherpa}} hyperparameter optimization library~\cite{SherpaOptimization}.
Additional details about the hyperparameters and their optimization are given in  Tables~\ref{tab:hyperparam_range_conv}, ~\ref{tab:hyperparam_range_dense} and~\ref{tab:hyperparam_range}.

\section{ Performance}
\label{sec:perf}

\begin{figure}[ht]
    \centering
    \includegraphics[width=\linewidth]{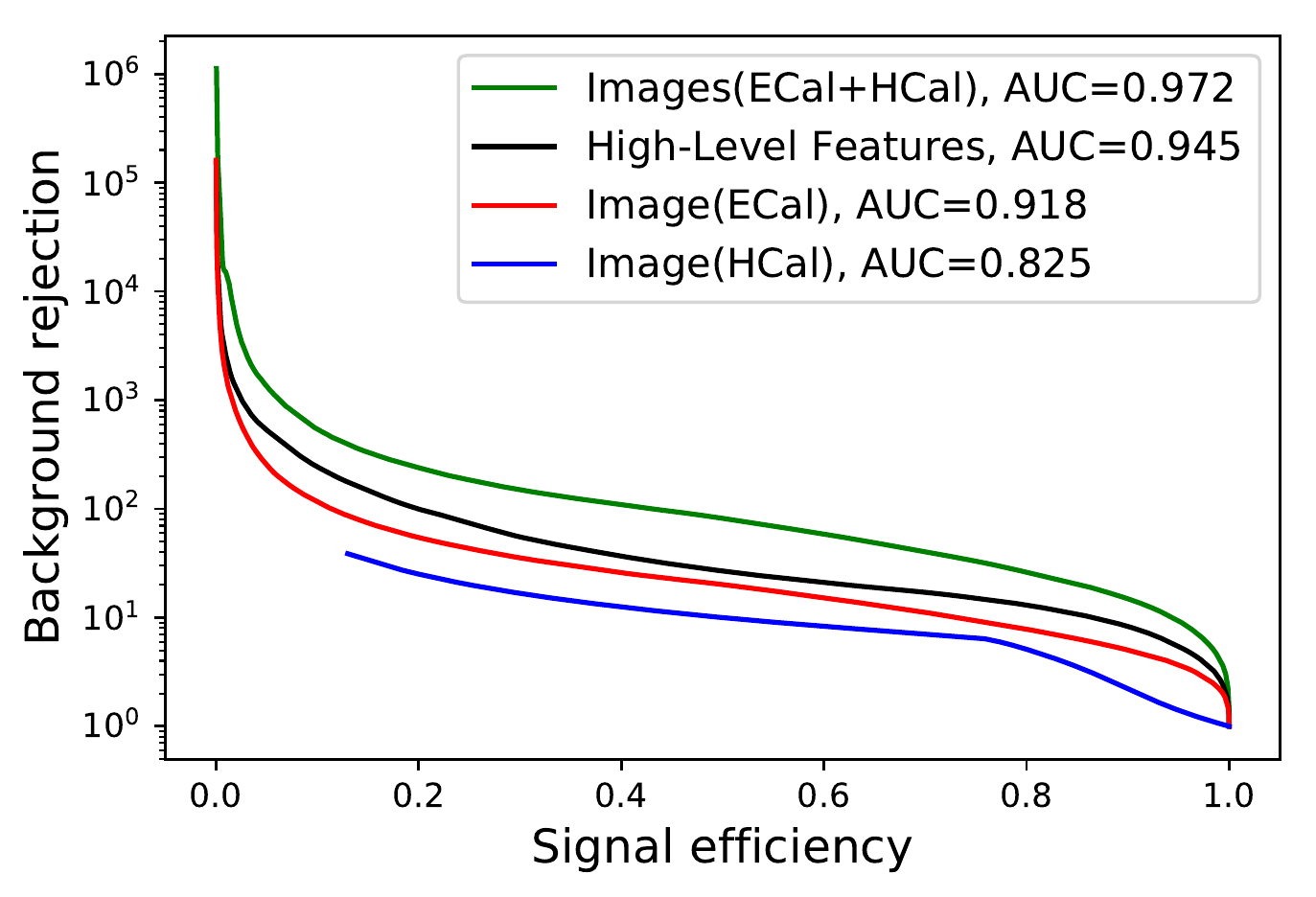}
    \caption{ Comparison of the  performance in electron identification for networks with varying sets of input features. Shown is the signal efficiency versus background rejection, and the AUC, for networks which use the existing set of expert high-level features (see text for details), networks which use HCal or ECal images, or both.}
    \label{fig:roc}
\end{figure}

Initial studies indicated that having images that reflect both $E$ and $E_\textrm{T}$ provided no performance boost, so only results with $E_\textrm{T}$-based images are shown here and used for further studies. A comparison of the performance of the image networks and the seven standard high-level features ($R_{\textrm{had}}$,  $\omega_{\eta 2}$,  $R_\phi$,  $R_\eta$, $\sigma_{\eta\eta}$, \mbox{Iso($\Delta R < 0.3$)}, \mbox{Iso($\Delta R < 0.4$)}) is shown in Fig.~\ref{fig:roc} and described in Table~\ref{tab:auc}.  

Networks combining the standard high-level features (AUC of 0.945) do not match the performance of a network which analyzes the lower-level data expressed as images (0.972), indicating that the images contain additional, untapped information relevant to the identification of electrons. This is not unexpected, and is in line with similar results for jet substructure or flavor tagging~\cite{Guest:2016iqz,Baldi:2016fql}.   Networks which see only one of the ECal or HCal images but not both do not match this performance, supporting the intuition that both calorimeters contribute valuable information.  Adding the HL features to the CNN, however, gives an almost negligible boost in performance, suggesting that the CNN has succeeded in capturing the power of the HL features.


\begin{table}[]
    \centering
        \caption{ Electron classification power (AUC) for networks with various feature sets. Images refer to low-level pixel data. Standard features are the high-level (HL) features typically used ($R_{\textrm{had}}$,  $\omega_{\eta 2}$,  $R_\phi$,  $R_\eta$, $\sigma_{\eta\eta}$, \mbox{Iso($\Delta R < 0.3$)}, \mbox{Iso($\Delta R < 0.4$)}), as described in the text.  All AUC values have an uncertainty of $\pm$ 0.001 unless otherwise specified.}
    \begin{tabular}{c|c|c|c|c}
    \hline\hline
        \multicolumn{4}{c|}{Network Features}& AUC  \\
        \hline
        \multicolumn{2}{c|}{Images} & 7 Standard   &  \\
        ECal  & HCal                & HL Features & $M_{\textrm{jet}}$   \\
        \hline
        &     \checkmark  & & &  0.82 $\pm$ 0.02\\
        \checkmark & & & & 0.918 \\
        \checkmark &     \checkmark  & & &  0.972\\
         \checkmark &     \checkmark  & \checkmark & &  0.973\\
                && \checkmark & & 0.945\\
                                && \checkmark & \checkmark & 0.956 \\
             \hline\hline
    \end{tabular}
    \label{tab:auc}
\end{table}

\section{ Bridging the gap}
\label{sec:gap}

The performance of the deep CNN reveals that there is information in the low-level image that is not captured by the suite of existing high-level features.  The goal, however, is not to replace the suite of features with an image-based network whose decisions are opaque to us and may not align with real physical principles. Instead, our aim is to identify new high-level features which bridge the gap between the existing performance and the superior performance of the CNN.

We note that the design of the high-level features focuses on highlighting the characteristics of the signal electrons, localized energy depositions primarily in the ECal without significant structure. The background, however, is due to jets, which potentially can exhibit a rich structure and comprise a mixture of jets from  gluons, light quarks, and heavy quarks. Each parton may produce jets with a distinct structure and varying probability to mimic electrons.   We hypothesize that features which are sensitive to the structure of the jet, or subclasses of jets, may provide additional discrimination power.

We first consider the powerful feature of jet mass, $M_\textrm{jet}$,  which is not often applied to electron identification, but has a distinct marginal distribution for electrons and jets, see Fig.~\ref{fig:hl}. Including it in a network of HL features provides a small but distinct boost in performance, see Table~\ref{tab:auc}, indicating that it contains useful information for this classification task not duplicated by the standard seven HL features.   This encourages us to explore further the space of jet observables as a way to understand the source of additional classification power of the CNN.  

\subsection{Set of Observables}

One could in principle consider an infinite number of jet observables. To organize our search,  we use the Energy Flow Polynomials (EFPs)~\cite{Komiske:2017aww}, a large (formally infinite) set of parameterized engineered functions, inspired by previous work on energy correlation functions \cite{Larkoski:2013eh}, which sum over the contents of the cells scaled by relative angular distances.

These parametric sums are described as the set of all isomorphic multigraphs where:
\begin{align}
	\text{each node} &\Rightarrow \sum_{i = 1}^N z_i, \label{eq:EFP_node}  \\
	\text{each $k$-fold edge} &\Rightarrow \left(\theta_{ij}\right)^k \label{eq:EFP_edge} . 
\end{align}

The observable corresponding to each graph can be modified with parameters $(\kappa, \beta)$, where
\begin{align}
	(z_i)^\kappa &= \left(\frac{p_{\textrm{T}i}}{\sum_j p_{\textrm{T}j}} \right)^\kappa, \label{eq:EFP_z} \\
	\theta^\beta_{ij} &= \left(\Delta \eta_{ij}^2 + \Delta \phi_{ij}^2 \right)^{\beta/2}. \label{eq:EFP_theta}
\end{align}

Here, $p_{\textrm{T}i}$ is the transverse momentum of cell $i$, and $\eta_{ij}$ ($\phi_{ij}$) is pseudorapidity (azimuth) difference between cells $i$ and $j$. 
The original IRC-safe EFPs require $\kappa = 1$, however we consider examples with $\kappa \not= 1$ to explore a broader space of observables. 
Also, note that $\kappa > 0$ generically corresponds to IR-safe but C-unsafe observables.\footnote{For $\kappa < 0$, empty cells are omitted from the sum.}  

 In principle, the space is complete, such that any jet observable can be described by one or more EFPs of some degree; in practice, the space is infinite and only a finite subset can be explored. We consider EFPs up to degree $d=7$ and with $\beta$ values of $\left[ \frac{1}{2},1,2\right]$ and $\kappa$ values of $\left[ -1,0,1,2\right]$.  We consider each graph as applied to the ECal or the HCal separately, effectively doubling the number of graphs\footnote{We also explored a version where ECal and HCal information were used simultaneously by each graph, but found no improvement.}.

\subsection{Searching for Observables}

Rather than conduct a brute-force search of this large space, we aim to leverage the success of the CNN and find observables which mimic its decisions.  We follow the black-box guided algorithm of Ref.~\cite{renn}, which isolates the portion of the input space where the CNN and existing HL features disagree and searches for a new observable that matches the decisions of the CNN algorithm in that subspace. 

The subspace is defined as   input pairs $(x,x')$ that have a different relative ordering between the CNN and the network of $n$ HL features (HLN$_n$). Mathematically, we express this using the {\it decision ordering} (DO) 

\begin{equation}
\label{eq:DO_def}
	\textrm{DO}[f,g](x,x') = \Theta \Big( \big( f(x) - f(x') \big) \big( g(x) - g(x') \big) \Big),
\end{equation}

\noindent
where $f(x)$ and $g(x)$ are classification functions such as the CNN or the HLN$_n$, such that DO$=0$ corresponds to inverted ordering and DO$=1$ corresponds to the same ordering. The focus of our investigation are the set of pairs  $X_n$ where the two classifiers disagree, defined as

\begin{equation}
	X_n = \Big\{ (x,x') \, \Big| \, \textrm{DO}\big[\textrm{CNN}, \textrm{HL}_n \big](x,x') = 0 \Big\}.
\end{equation}

As prescribed in Ref.~\cite{renn}, we scan the space of EFPs to find the observable that has the highest average decision ordering (ADO) with the CNN when averaged over the disordered pairs $X_n$.  The selected EFP is then incorporated into the new network of HL features, HLN$_{n+1}$, and the process is repeated until the ADO or AUC plateaus.

For all HLN$_n$ used in this search, models were trained with {\sc{Keras}}~\cite{chollet2015keras} using {\sc{Tensorflow}}~\cite{tensorflow2015} as the backend. Each model was built as a fully connected neural network of simple one dimensional input features and a single logistic unit output. These networks consisted of 3 hidden layers, each with 50 rectified linear units, separated by 2 dropout layers using a dropout value of 0.25 and trained with a batch size of 128. The {\sc{Adam}} optimizer~\cite{Adam2014} was used with learning rate of 0.001 and initialized with glorot normal weights.

\subsection{IRC safe observables}

We begin our search by considering only the observables which are IRC safe, with $\kappa=1$.  Beginning with the seven HL features, the first  graph selected is:

\[ \gxPlot{0.075}{7HL_irc_safe}{1} = \sum^N_{a,b=1} z_a z_b \theta_{ab}^{\frac{1}{2}} \]

\noindent with $ \beta=\frac{1}{2}$. This graph has an ADO of 0.802 with the CNN over the input subspace where the CNN disagrees with the seven HL, suggesting that it is well aligned with the CNN's strategy. Adding it to the seven HL features achieves an AUC of $0.970\pm 0.001$, very nearly closing the gap with the CNN performance of 0.972.   This graph is very closely related to jet mass, a pairwise sum over cells which folds in angular separation, but more closely resembles the Les Houches Angularity variable~\cite{Gras:2017jty}, which similarly is sensitive to the distribution of energy away from the center, though with a smaller power of the angularity than jet mass, which suggests that it enhances small angles. Additional scans do not identify EFP observables with a useful ADO and do not contribute to the AUC.

If instead, we begin with the seven HL features as well as the jet mass,  the procedure selects two graphs: 

\[ \gxPlot{0.05}{7HL_mass_irc_safe}{1} = \sum^N_{a\cdots h=1}  z_a z_b z_c z_d z_e z_f z_g z_h \theta_{ab} \theta_{ac} \theta_{ad} \theta_{ae} \theta_{af} \theta_{ag} \theta_{ah}\]

\noindent
and

\[ \gxPlot{0.05}{7HL_mass_irc_safe}{2} = \sum^N_{a,b,c=1}  z_a z_b z_c \theta_{ab}^{\frac{1}{2}} \theta_{bc}^{\frac{1}{2}} \theta_{ac}^{\frac{1}{2}} \]

\begin{figure}[h!]
	\centering
    \begin{tikzpicture}
    \node(a){
	\includegraphics[width=0.3\textwidth]{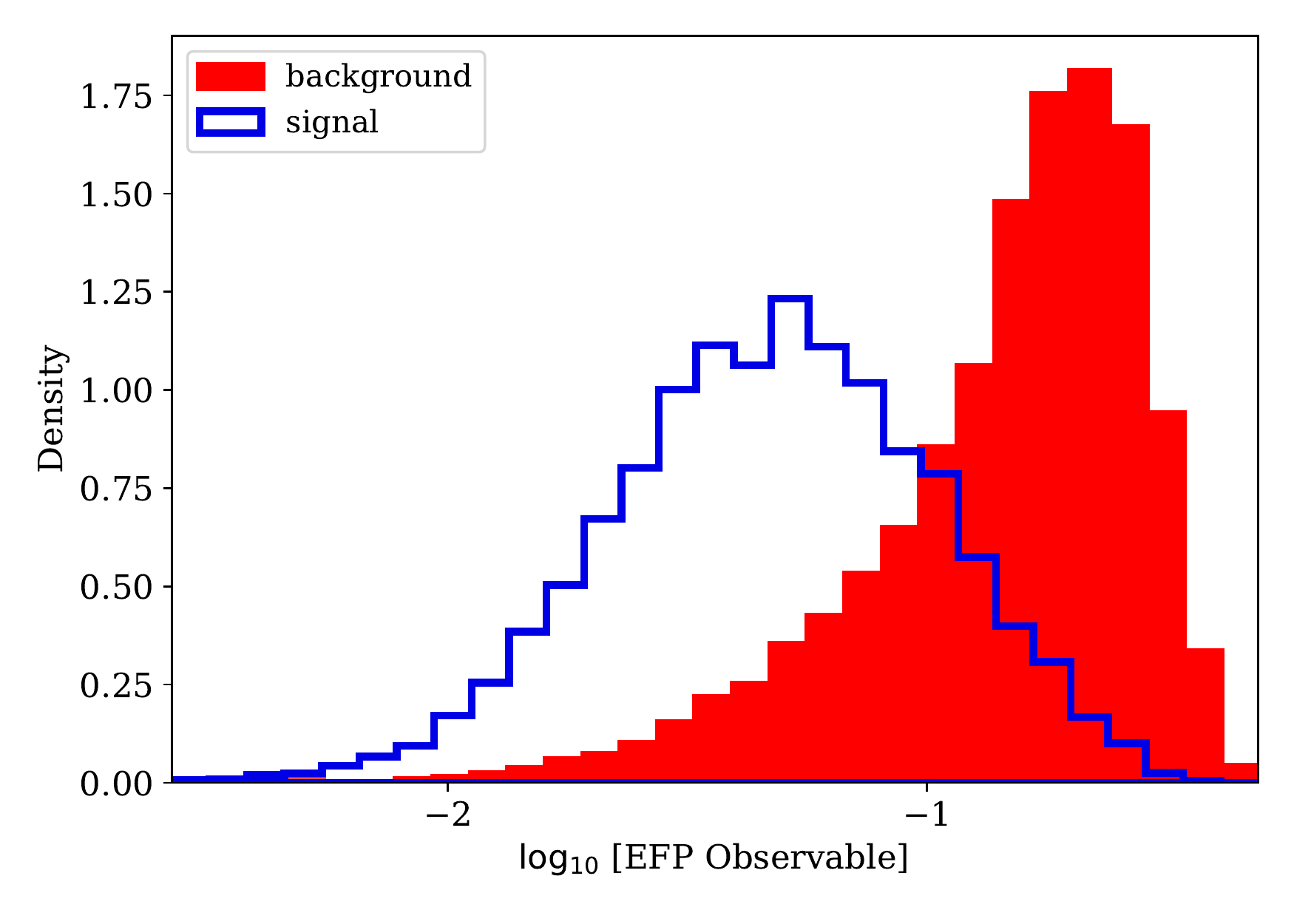}};
    \node at (a.north east)
    [
    anchor=center,
    xshift=-42mm,
    yshift=-12mm
    ]
    {
    \setlength{\fboxrule}{0.02pt}%
    $\gxPlot{0.035}{7HL_irc_safe}{1}$
    };
    \node at (a.north east)
    [
    anchor=center,
    xshift=-40mm,
    yshift=-17mm
    ]
    {
    \setlength{\fboxrule}{0.01pt}%
    \scalebox{0.6}{$\kappa=1,\beta=\frac{1}{2}$}
    };
    \end{tikzpicture}
     \begin{tikzpicture}
    \node(a){
	\includegraphics[width=0.3\textwidth]{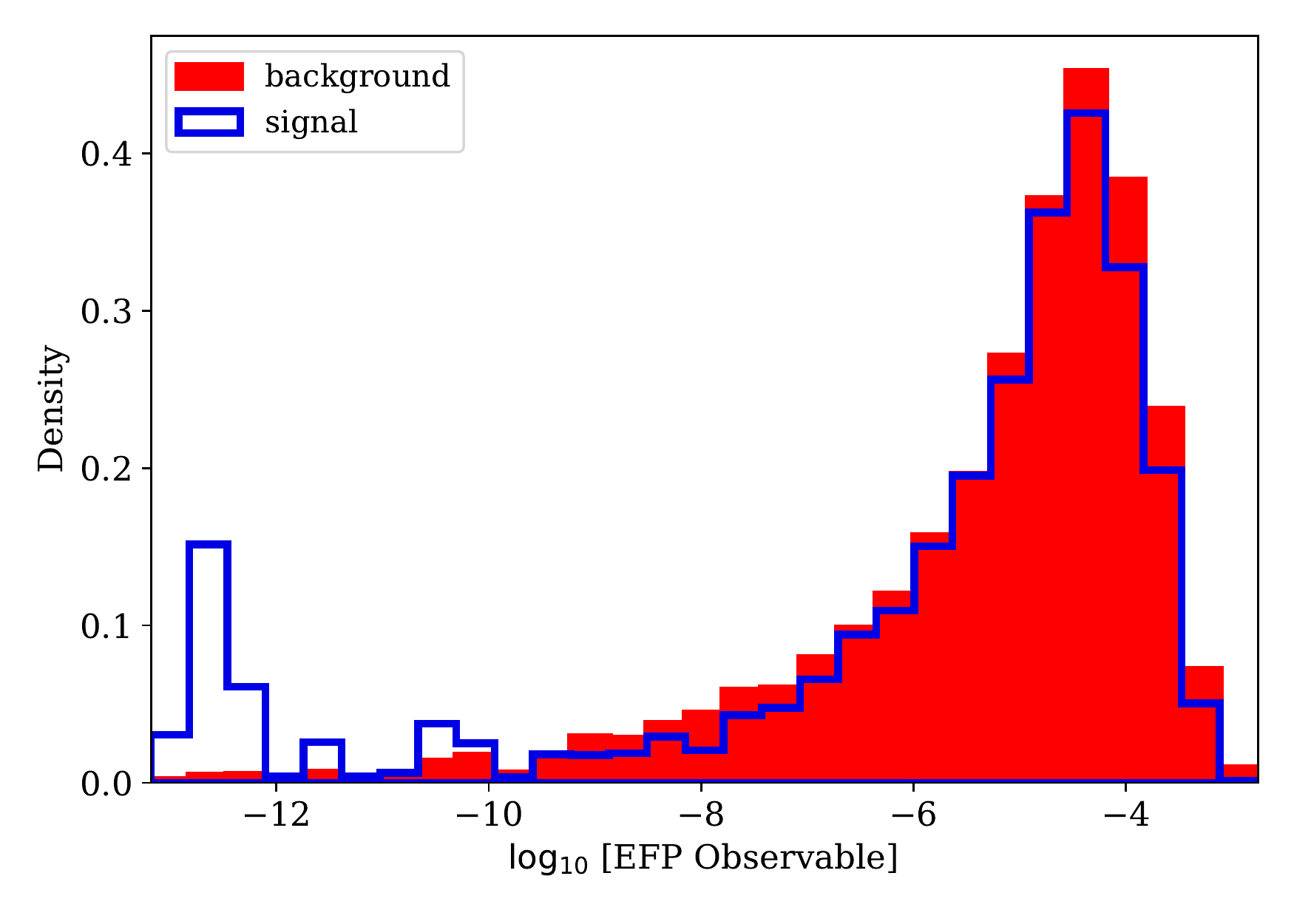}};
    \node at (a.north east)
    [
    anchor=center,
    xshift=-42mm,
    yshift=-12mm
    ]
    {
    \setlength{\fboxrule}{0.02pt}%
    $\gxPlot{0.035}{7HL_mass_irc_safe}{1}$
    };
    \node at (a.north east)
    [
    anchor=center,
    xshift=-40mm,
    yshift=-17mm
    ]
    {
    \setlength{\fboxrule}{0.01pt}%
    \scalebox{0.6}{$\kappa=1,\beta=1$}
    };
    \end{tikzpicture}
      \begin{tikzpicture}
    \node(a){
	\includegraphics[width=0.3\textwidth]{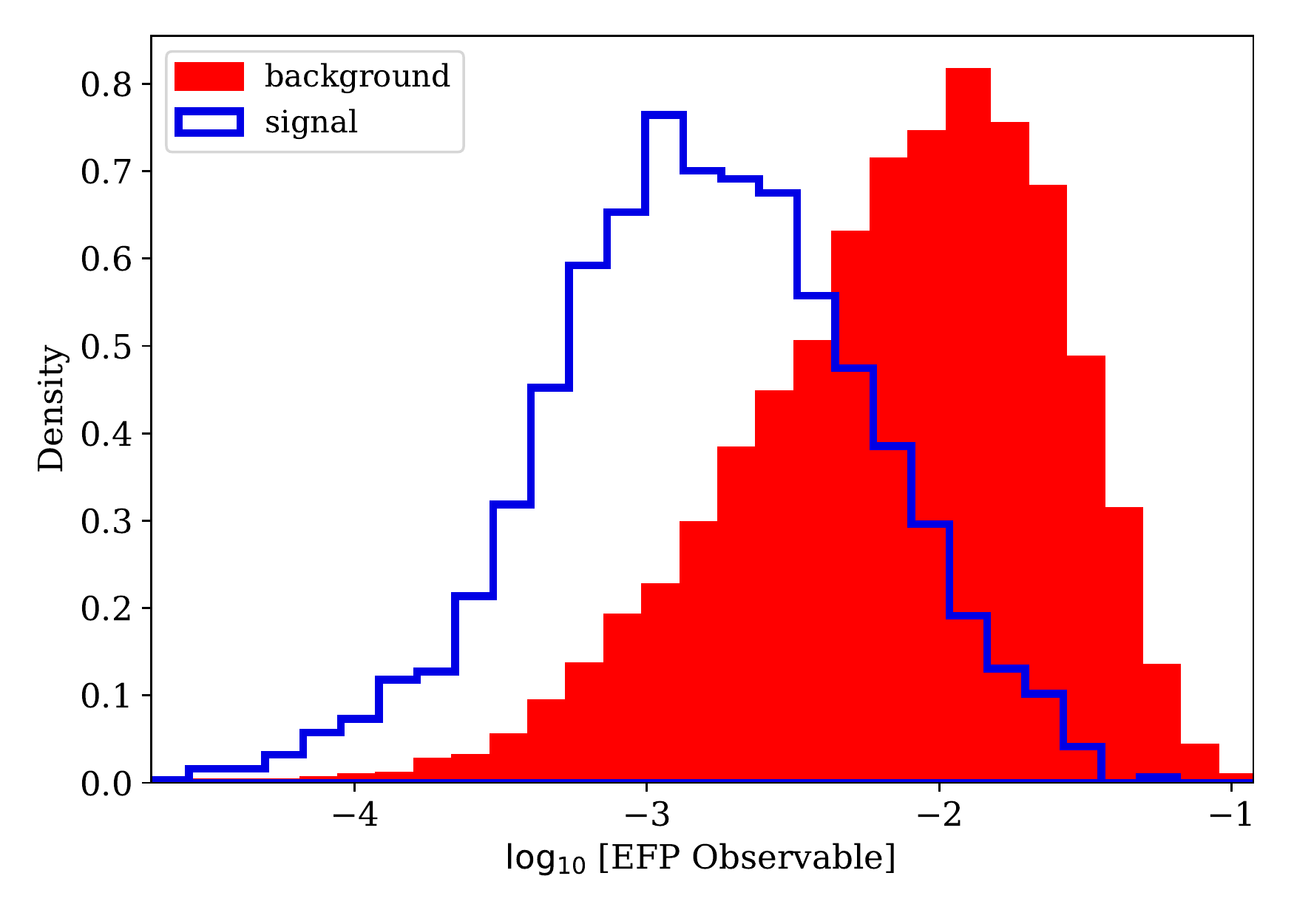}};
    \node at (a.north east)
    [
    anchor=center,
    xshift=-42mm,
    yshift=-12mm
    ]
    {
    \setlength{\fboxrule}{0.02pt}%
    $\gxPlot{0.035}{7HL_mass_irc_safe}{2}$
    };
    \node at (a.north east)
    [
    anchor=center,
    xshift=-40mm,
    yshift=-17mm
    ]
    {
    \setlength{\fboxrule}{0.01pt}%
    \scalebox{0.6}{$\kappa=1,\beta=\frac{1}{2}$}
    };
    \end{tikzpicture}
	\caption{$\log_{10}$ distributions of the selected IRC-safe EFPs as chosen by the black-box guided strategy, for signal electrons and background jets.}
	\label{fig:efp_7hl_safe}
\end{figure}
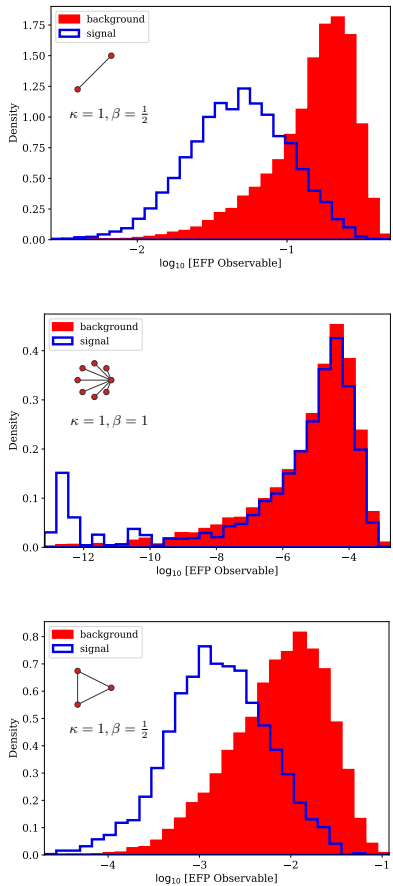

\noindent
When combined with the seven HL features and $M_{\textrm{jet}}$, this set of ten observables achieves an AUC of $0.971 \pm 0.001$, almost matching the CNN performance. Distributions of these observables for signal and background samples are shown in Fig.~\ref{fig:efp_7hl_safe}.

\subsection{Broader Scan}

In this Section, we present a  scan of a larger set of EFPs, including values of $\kappa$ which lead to IRC unsafe observables, $\kappa = [-1,0,1,2]$.

Beginning from the seven standard HL features, the first pass selects a simple observable:

\[ \gxPlot{0.075}{7HL}{1} = \sum^N_{a=1} z_a^2 \]

\noindent 
with no angular terms at all, but $\kappa=2$. This is known in the jet substructure literature as $p_\textrm{T}^D$~\cite{Pandolfi:2012ima,Chatrchyan:2012sn} and was originally developed to help distinguish between quark and gluon jets. When combined with the other seven HL features,  this observable also reaches a performance of $0.970 \pm 0.001$.  Further scans do not lead to statistically significant improvements in AUC.

\begin{figure}[h!]
	\centering
	\begin{tikzpicture}
    \node(a){
	\includegraphics[width=0.3\textwidth]{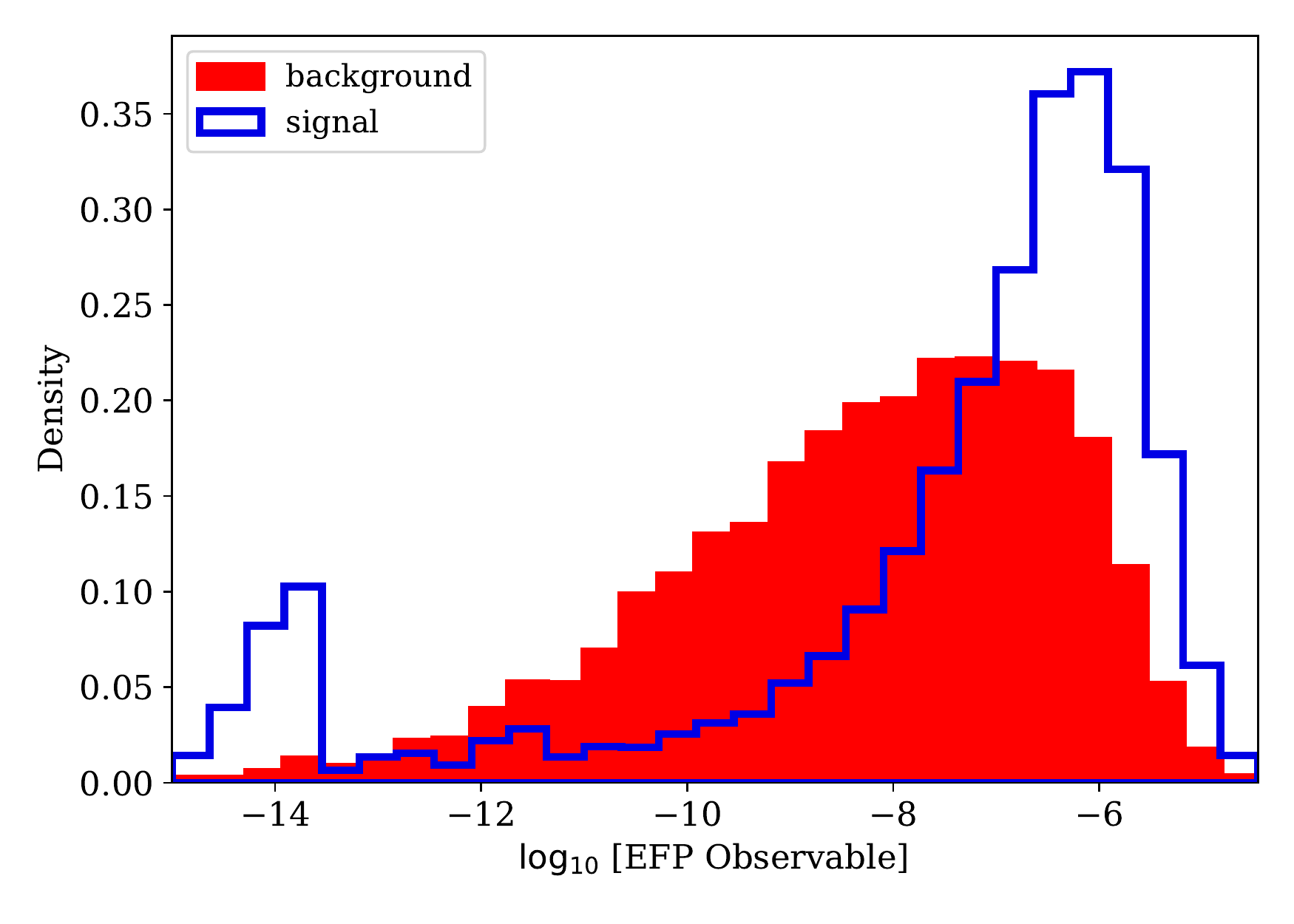}};
    \node at (a.north east)
    [
    anchor=center,
    xshift=-42mm,
    yshift=-12mm
    ]
    {
    \setlength{\fboxrule}{0.02pt}%
    $\gxPlot{0.035}{7HL_mass_irc_safe}{1}$
    };
    \node at (a.north east)
    [
    anchor=center,
    xshift=-40mm,
    yshift=-17mm
    ]
    {
    \setlength{\fboxrule}{0.01pt}%
    \scalebox{0.6}{$\kappa=2,\beta=1$}
    };
    \end{tikzpicture}
	\begin{tikzpicture}
    \node(a){
	\includegraphics[width=0.3\textwidth]{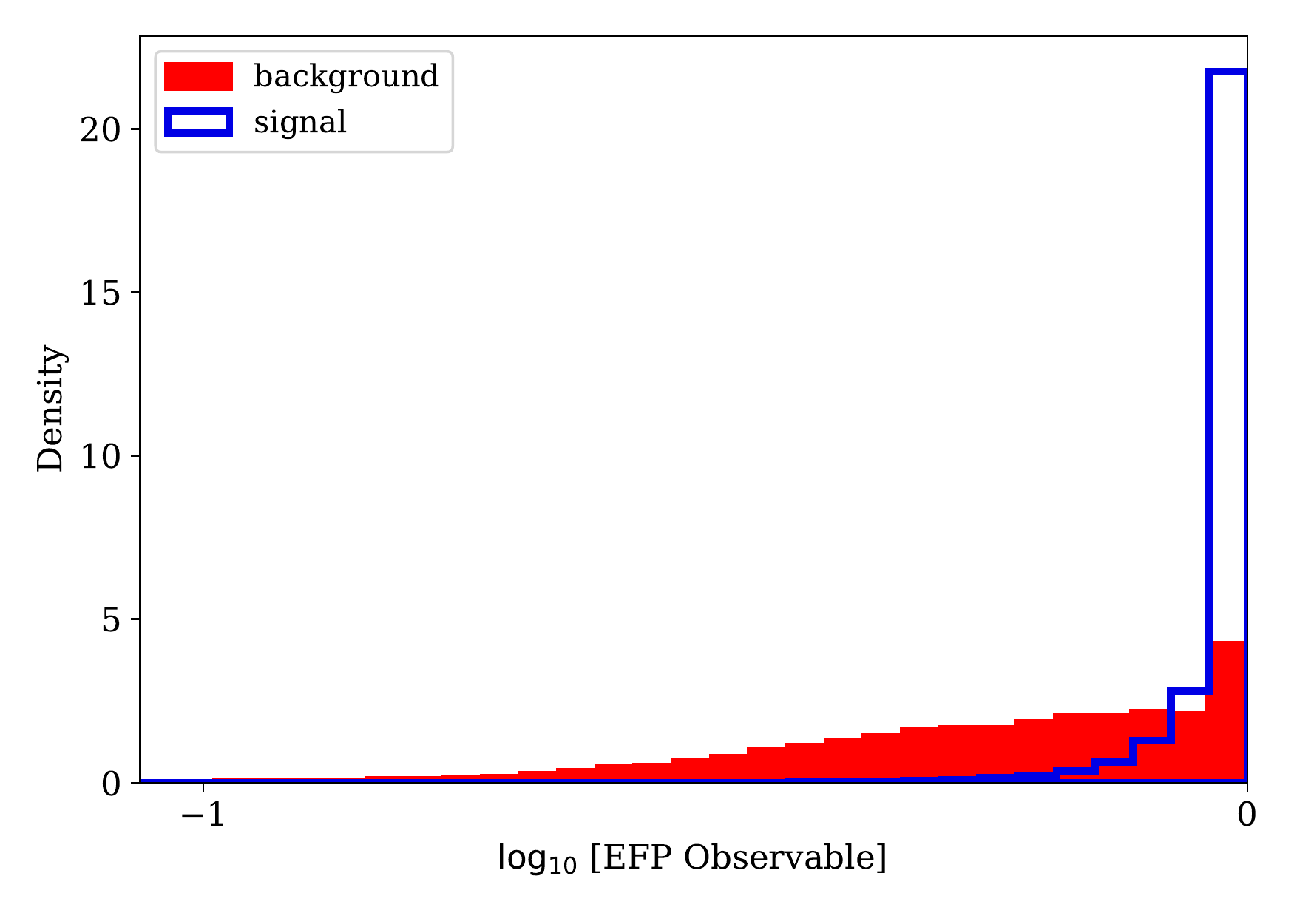}};
    \node at (a.north east)
    [
    anchor=center,
    xshift=-42mm,
    yshift=-12mm
    ]
    {
    \setlength{\fboxrule}{0.02pt}%
    $\gxPlot{0.035}{7HL}{1}$
    };
    \node at (a.north east)
    [
    anchor=center,
    xshift=-40mm,
    yshift=-17mm
    ]
    {
    \setlength{\fboxrule}{0.01pt}%
    \scalebox{0.6}{$\kappa=2$}
    };
    \end{tikzpicture}
	\caption{$\log_{10}$ distributions of the selected EFPs as chosen by the black-box guided strategy, regardless of IRC safety, for signal electrons and background jets.}
	\label{fig:efp_7hl}
\end{figure}
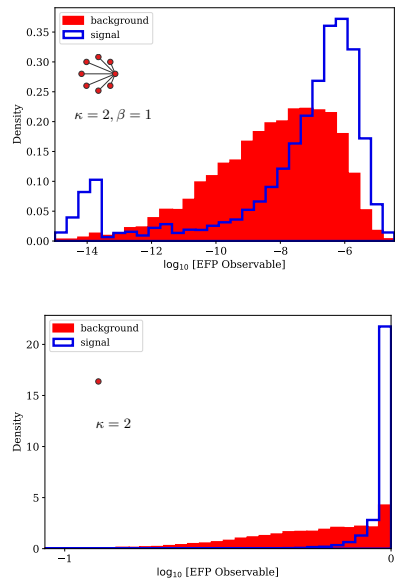

If instead, we begin from the seven standard HL features and $M_\textrm{jet}$, we find $\gxPlot{0.025}{7HL_mass_irc_safe}{1}$, this time with $\kappa=2$ as well as the simpler $p_\textrm{T}^D$.  Distributions of these two IRC unsafe EFP observables for signal and background are shown in Fig.~\ref{fig:efp_7hl}. Together with the seven HL and $M_\textrm{jet}$, these 10 observables reach a performance of $0.971 \pm 0.001$.  Further scans do not lead to statistically significant improvements in AUC.

See Table~\ref{tab:perf_summary} for a summary of the additional observables needed to reach the performance of $\approx 0.97$ in each case.

\begin{table}[t]
\centering
\caption{Summary of the performance of various networks considered. Uncertainty in the AUC value is $\pm 0.001$, estimated using bootstrapping.}
\label{tab:perf_summary}
\begin{tabular}{ll|l|l|r}
\hline\hline
Base & &\multicolumn{2}{c}{Additions ($\kappa,\beta$) } &  (AUC) \\
\hline
7HL & & & & 0.945\\
7HL & $+M_{\textrm{jet}}$ & & & 0.956 \\
7HL & & $\gxPlot{0.02}{7HL_irc_safe}{1}$ ($1,\frac{1}{2}$)  &  & 0.970 \\
7HL & $+M_{\textrm{jet}}$ & $\gxPlot{0.025}{7HL_mass_irc_safe}{1}$ ($1,1$) &  $\gxPlot{0.02}{7HL_mass_irc_safe}{2}$ ($1,\frac{1}{2}$) & 0.971 \\
7HL & & $\gxPlot{0.025}{7HL}{1}$ ($2,-$) & & 0.970 \\
7HL & $+M_{\textrm{jet}}$ &   $\gxPlot{0.025}{7HL_mass_irc_safe}{1}$ ($2,1$) &  $\gxPlot{0.025}{7HL}{1}$ ($2,-$) & 0.971 \\
\hline
CNN & & & & 0.972\\
\hline
\hline
\end{tabular}
\end{table}

\section{ Discussion}
\label{sec:disc}

Our deep neural networks indicate that low-level calorimeter data represented as images contains information useful for the task of electron identification that is not captured by the standard set of high-level features as implemented here.

A guided search~\cite{renn} through the EFP space identified two EFP observables calculated on the ECal cells which mimic the CNN strategy and bridge the gap. Observables on the HCal information were not helpful to the classification task. The first,

\[ \gxPlot{0.075}{7HL_irc_safe}{1} = \sum^N_{a,b=1} z_a z_b \theta_{ab}^{\frac{1}{4}} \]

\noindent is closely related to the Les Houches Angularity~\cite{Gras:2017jty}, and confirms our suspicion that the non-trivial structure of the background object provides a useful handle for classification.  The second observable,  $p_\textrm{T}^D$~\cite{Pandolfi:2012ima,Chatrchyan:2012sn},

\[ \gxPlot{0.075}{7HL}{1} = \sum^N_{a=1} z_a^2 \]

\noindent with $\kappa=2$ is not IRC safe, and was originally developed to help distinguish between quark and gluon jets. It effectively counts the number of hard particles, which is sensitive to the amount of color charge, where electrons and jets are clearly distinct.

Both Les Houches Angularity and $p_\textrm{T}^D$ display power to separate electrons from the jet backgrounds, by exploiting the structure and nature of the jet energy deposits.  While the precise performance obtained here may depend at some level on the fidelity of the simulation used and the resulting limitations on the implementation of state-of-the-art high-level features, these results strongly suggest that these observables be directly studied in experimental contexts where more realistic simulation tools are available, or directly in data samples, using weakly supervised learning~\cite{Dery:2017fap}.

More broadly, the existence of a gap between the performance of state-of-the-art high-level features and CNN represents an opportunity to gather additional power in the battle to suppress lepton backgrounds. Rather than employing black-box CNNs directly, we have demonstrated the power of using them to identify the relevant observables from a large list of physically interpretable options.  This allows the physicist to understand the nature of the information being used and to assess its systematic uncertainty.

 Any boost in electron identification performance is extremely valuable to searches at the LHC, especially those with multiple leptons, where event-level efficiencies depend sensitively on object-level efficiencies.

All code and data used in this project is available at: https://github.com/TDHTTTT/EID, as well as through the UCI Machine Learning in Physics web portal at:
http://mlphysics.ics.uci.edu/.

\section{ Acknowledgments}

The authors thank Jesse Thaler, Ian Moult and Tim Tait for helpful discussions. We wish to acknowledge a hardware grant from NVIDIA. This material is based upon work supported by the National Science Foundation under grant number 1633631. The work of JC and PB is in part supported by grants NSF 1839429 and NSF 
NRT 1633631 to PB. The work of JNH is in part supported by grants DE-SC0009920, DGE-1633631, and DGE-1839285.  TT wants to thank Undergraduate Research Opportunities Program at UCI for grant number 02399s1. 

\begin{center}
\line(1,0){100}
\end{center}

\clearpage

\clearpage
\section{ Neural Network Hyperparameters and Architecture}

\begin{table}[h]
    \centering
        \caption{Hyperparameter ranges for bayesian optimization of convolutional networks}
            \label{tab:hyperparam_range_conv}
    \begin{tabular}{c|c}
    \hline\hline
        Parameter &  Range\\
        \hline
          Num. of conv. blocks & [1, 4] \\
          Num. of filters & [8, 128]  \\
          Num. of dense layers & [1, 3] \\
          Num. of hidden units & [1, 200]\\
          Learning rate & [0.0001, 0.01] \\
          Dropout & [0.0, 0.5] \\
             \hline\hline
    \end{tabular}
\end{table}

\begin{table}[h]
    \centering
        \caption{Hyperparameter ranges for bayesian optimization of fully connected networks}
    \label{tab:hyperparam_range_dense}
    \begin{tabular}{c|c}
    \hline\hline
        Parameter &  Range  \\
        \hline
          Num. of dense layers & [1, 8] \\
          Num. of hidden units & [1, 200]  \\
          Learning rate & [0.0001, 0.01] \\
          Dropout & [0.0, 0.5] \\
             \hline\hline
    \end{tabular}
\end{table}

\begin{table}[h]
    \centering
        \caption{Best hyperparameters found per model.}
    \begin{tabular}{c|c|c|c|c|c|c}
    \hline\hline
       features & conv. & filters & dense & hidden & LR & DP  \\
        \hline
          ECal & 3 & 117 & 2 & 160 & 0.0001 & 0.0 \\
          Hcal & 2 & 27 & 2 & 84 & 0.01 & 0.5 \\
          Ecal+HCal & 3 & 47 & 2 & 146 & 0.0001 & 0.0 \\
          HL & - & - & 5 & 149 & 0.001 & 0.0019 \\
             \hline\hline
    \end{tabular}
    \label{tab:hyperparam_range}
\end{table}

\begin{figure}[ht]
    \centering
    \includegraphics[width=0.8\linewidth]{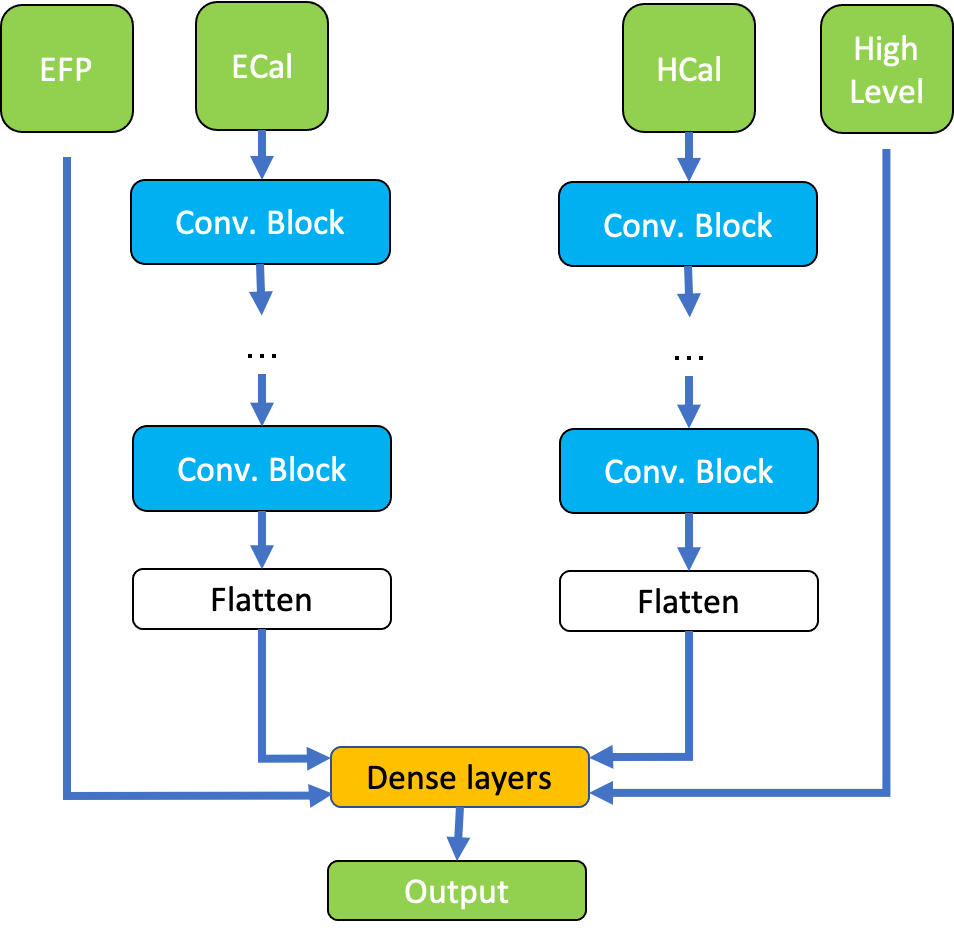}
    \caption{Diagram of the architecture of the convolutional neural network.}
    \label{fig:architecture}
\end{figure}

\begin{figure}[ht]
    \centering
    \includegraphics[width=0.8\linewidth]{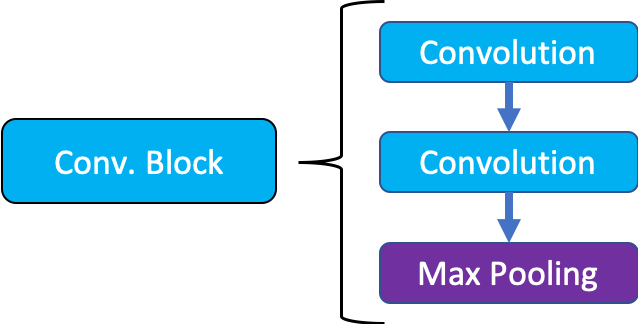}
    \caption{Diagram of convolutional block appearing in network architecture, see Fig~\ref{fig:architecture}.}
    \label{fig:convolutionalblock}
\end{figure}

\clearpage

\bibliography{electron}

\end{document}